\documentclass[lettersize,journal]{IEEEtran}
\usepackage{amsmath,amsfonts}
\usepackage{algorithm}
\usepackage{array}
\usepackage[caption=false,font=normalsize,labelfont=sf,textfont=sf]{subfig}
\usepackage{textcomp}
\usepackage{stfloats}
\usepackage{url}
\usepackage{verbatim}
\usepackage{graphicx}
\usepackage{cite}
\usepackage{algpseudocode,cases,multirow,booktabs}

\begin{document}

\title{Neural Network for Blind Unmixing: a novel MatrixConv Unmixing (MCU) Approach}

\author{Chao~Zhou,~\IEEEmembership{Student Member,~IEEE,}
        Wei~Pu,~\IEEEmembership{Member,~IEEE,}
        and~Miguel~Rodrigues,~\IEEEmembership{Fellow,~IEEE}
\thanks{C. Zhou and M. R.D.~Rodrigues are with the Department of Electronic and Electrical Engineering, University College London, London, WC1E 6BT, U.K. Wei Pu is with the Department of Information and Communication Engineering, University of Electronic Science and Technology of China, Chengdu, Sichuan, 611731, China. (e-mail: chao.zhou.18@ucl.ac.uk; pwuestc@163.com; m.rodrigues@ucl.ac.uk). Corresponding Author: Wei~Pu.}
\thanks{Manuscript received April 19, 2021; revised August 16, 2021.}}

\markboth{Journal of \LaTeX\ Class Files,~Vol.~14, No.~8, August~2021}%
{Shell \MakeLowercase{\textit{et al.}}: A Sample Article Using IEEEtran.cls for IEEE Journals}


\maketitle

\begin{abstract}
Hyperspectral image (HSI) unmixing is a challenging research problem that tries to identify the constituent components, known as endmembers, and their corresponding proportions, known as abundances, in the scene by analysing images captured by hyperspectral cameras. Recently, many deep learning based unmixing approaches have been proposed with the surge of machine learning techniques, especially convolutional neural networks (CNN). However, these methods face two notable challenges: 1. They frequently yield results lacking physical significance, such as signatures corresponding to unknown or non-existent materials. 2. CNNs, as general-purpose network structures, are not explicitly tailored for unmixing tasks. In response to these concerns, our work draws inspiration from double deep image prior (DIP) techniques and algorithm unrolling, presenting a novel network structure that effectively addresses both issues. Specifically, we first propose a MatrixConv Unmixing (MCU) approach for endmember and abundance estimation, respectively, which can be solved via certain iterative solvers. We then unroll these solvers to build two sub-networks, endmember estimation DIP (UEDIP) and abundance estimation DIP (UADIP), to generate the estimation of endmember and abundance, respectively. The overall network is constructed by assembling these two sub-networks. In order to generate meaningful unmixing results, we also propose a composite loss function. To further improve the unmixing quality, we also add explicitly a regularizer for endmember and abundance estimation, respectively. The proposed methods are tested for effectiveness on both synthetic and real datasets.
\end{abstract}


\section{Introduction}
\IEEEPARstart{H}{yperspectral} image sensors can capture very rich spectral reflectance in hundreds or thousands of spectral channels in a scene~\cite{bioucas2013hyperspectral}. Because of richer spectral information compared to a normal RGB image, one can identify materials in the scene more accurately from an HSI image~\cite{Hong2021}. However, the captured spectral reflectance is often a mixture of the spectral signatures of the materials present in the scene. Thus, it requires methods to quantitatively unmix the observed spectral reflectances onto the spectral signatures of its constituents, also known as endmembers, and the corresponding proportions, also known as abundances~\cite{keshava2002spectral}. One of the most popular unmixing models is the linear mixture model (LMM), which presumes the observed spectral reflectance is a linear combination of the endmembers' signatures weighted by the corresponding abundances. Under LMM, the HSI unmixing problem involves two tasks~\cite{camps2011remote}: (a) \textit{endmember estimation (EE)} and (b) \textit{abundance estimation (AE)}.

Most traditional model-based unmixing methods solve EE and AE separately. For the EE task, due to the linearity of LMM, the observed reflectances are often assumed to be embedded in a simplex with endmembers at its vertices, inspiring approaches like vertex component analysis (VCA)~\cite{Nascimento2005} and simplex volume maximization (SiVM)~\cite{Heylen2011}. In the AE task, endmembers are assumed known or estimated priorly, reducing the task to a least square problem solvable by a fully constraint least square (FCLS)~\cite{Heinz2001} solver. When endmembers are known as a rich spectral library, the sparsity prior can be introduced, leading to the linear sparse regression problem~\cite{Bioucas-Dias2010} addressed by techniques like sparse unmixing by variable splitting and augmented Lagrangian (SunSAL)~\cite{Bioucas-Dias2010} and collaborative SUnSAL (CLSunSAL)~\cite{Iordache2014}. Blind Unmixing (BU) methods, like Nonnegative Matrix Factorization (NMF)~\cite{Miao2007}, address both EE and AE simultaneously. NMF treats HSI unmixing as a matrix factorization problem, incorporating priors such as non-negativity and total-variation (TV) constraints. However, these traditional model-based approaches are often computationally complex and depend on carefully crafted priors.

With the prosperity of machine learning techniques, many learning-based approaches have emerged to deal with HSI unmixing problems~\cite{Zhao2021,Jin2021,Min2021,Xiong2021,Shahid2021,Zhou2021}. Typically, these methods can be categorised into two classes: supervised and unsupervised methods. In the former class~\cite{Zhou2021,Licciardi2011,Zhang2018}, the learning models, which are usually deep neural networks (DNN) or convolutional neural networks (CNN), assume access to a set of pairs of HSI observations and corresponding abundances. The models then learn a mapping function from the observations to the corresponding abundances via minimising a predefined loss function. These methods however need access to the true abundance. On the contrary, unsupervised methods~\cite{Palsson2017,Palsson2018,Qu2019,Ozkan2019,Ozkan2019b,Su2019,Borsoi2020} can estimate abundances and endmembers simultaneously given only access to HSI reflectances. These blind unmixing approaches generally adopt the so-called autoencoder structure that takes HSI reflectances as input and enforces the output to be reconstructions of the HSI reflectances. The endmembers are embedded in the weights of the decoder which is usually linear, and the abundances are usually given by the bottleneck of the autoencoder. However, most learning-based approaches, lacking proper guidance, may not assure the generation of physically meaningful unmixing results, potentially leading to signatures corresponding to unknown or non-existent materials~\cite{Hong2021}.

To mitigate the above issues, recently deep image prior (DIP)~\cite{DIP2018} techniques have been applied in HSI unmixing problems to guarantee meaningful unmixing results~\cite{UnDIP2021}. Typically, they estimate abundance by relying on existing unmixing methods to generate endmember estimations for training guidance. Therefore, these networks do not conduct endmember estimations themselves, making them susceptible to any errors present in the provided endmember guidance. Furthermore, it has been demonstrated~\cite{Zhou2022} that current DIP-based unmixing networks suffer from a drawback: the unmixing performance is constrained by the quality of the provided guidance. In simpler terms, with an existing endmember estimation, training a DIP network to estimate abundances, does not outperform the abundance estimation achieved using a straightforward FCLS solver. 

Additionally, traditional DIP techniques~\cite{DIP2018,UnDIP2021} often replace explicit regularizers with learned implicit ones. However, it has been suggested~\cite{DeepRED} that explicit regularizers enhance the performance of DIP. Furthermore, these unmixing networks have often utilized general-purpose DNN or CNN structures not tailored to hyperspectral image characteristics, potentially leading to sub-optimal performance in HSI tasks~\cite{paoletti2021adaptable}. 

The challenge remains in designing DIP-based unmixing frameworks that can achieve meaningful estimates of endmembers and abundances simultaneously and surpass the limitations imposed by provided guidance. Furthermore, there is a need to elucidate how the design of networks tailored for unmixing can be guided, considering the unique features of HSI images versus typical RGB images. Additionally, there is a necessity for an exploration into how existing DIP-based unmixing networks can derive benefits from the integration of explicit regularizers.

Algorithm unrolling techniques~\cite{gregor2010learning} have emerged as an mechanism to design problem-specific networks, starting from iterative solvers like the Iterative Shrinkage-Thresholding Algorithm (ISTA), interpreting iterative updates as neural network layers with learnable solver parameters. While some unmixing works~\cite{Qian2020,Zhou2021b} have demonstrated the effectiveness by unrolling iterative solvers, their potential is limited by reliance on linear mixing models without leveraging convolutional operators. Despite the established efficacy of convolutional operators in image processing~\cite{krizhevsky2017imagenet}, the method to integrate them into a network tailored for unmixing problems, rather than a general-purpose CNN, remains unclear.

In this work, we propose a novel approach for HSI blind unmixing, using a network architecture that leverages convolutional sparse coding (CSC) techniques. Specifically, we introduce a novel MatrixConv Unmixing (MCU) approach to capture global and local features through multiplication and convolution. We then employ the alternating direction method of multipliers (ADMM) solver to the approach. Two DIP architectures, UEDIP and UADIP, are constructed for endmember and abundance estimation, by unrolling the ADMM solver. Combining these forms our final \textbf{N}etwork for \textbf{B}lind-unmixing based on \textbf{A}DMM (NBA). NBA is trained using a composite loss function for physically meaningful estimations. Explicit regularizer by denoising (RED)~\cite{RED,DeepRED} are added to enhances endmember and abundance estimation. Our pioneering work incorporates CSC into the linear mixing model and proposes DIP network construction methods, paving the way for task-dependent DIP networks. Concretely, our main contributions are:
\begin{enumerate}
    \item Inspired by CSC techniques, we present a novel MatrixConv Unmixing (MCU) approach for the HSI unmixing task, which can be solved by ADMM solvers. By applying algorithm unrolling techniques to the ADMM solvers, two DIP networks, namely UEDIP and UADIP, are constructed for endmember and abundance estimation, respectively. Unlike generic DNN or CNN architectures, the proposed network architecture, which is tailored for unmixing tasks, can capture both global and local features through a combination of multiplication and convolution operations;
    \item we present a fully blind unmixing network by combining UEDIP and UADIP according to the LMM model. To make sure that the network can output meaningful unmixing results and achieve better performance than the guidance, we propose to train it with a composite loss function;
     \item We also add explicitly a RED regularizer for the estimation of abundance and endmember, respectively, to further improve the performance.
\end{enumerate}

This work is different form a recent conference paper~\cite{Zhou2022} as follows: (1) we propose a novel MatrixConv Unmixing (MCU) approach for EE and AE, respectively, and propose an iterative solver for each approach (section~\ref{sec: MCU-based HSI model}). (2) the EDIP (section~\ref{sec: Unfolding based EDIP}) and ADIP (section~\ref{sec: Unfolding based ADIP}) network structures are constructed by applying algorithm unrolling to the proposed iterative solvers. (3) we also add explicit regularizers to the estimation of endmembers and abundance, respectively, to improve the unmixing performance (section~\ref{sec: regularizer}).

\textit{Notations.}
In this work, $\ast$ denotes the convolutional operator and $\times$ denotes matrix multiplication. $\|\mathbf{X}\|_F=\sqrt{\sum_{i}\sum_{j}|x_{ij}|^2}$ denotes the Frobenius norm of matrix $\mathbf{X}$, $\|\mathbf{X}\|_{1}=\max_{j}\sum_{i}|x_{ij}|$ denotes the $\ell_{1}$ norm of matrix $\mathbf{X}$. $\mathbf{X}^T$ is the transform of matrix $\mathbf{X}$.

\section{Related work}
In this section, we introduce related works on HSI unmixing formulations, algorithm unrolling based unmixing methods and DIP based unmixing methods.

\subsection{HSI Blind Unmixing}
The linear mixing model (LMM) is one of the most popular unmixing models in hyperspectral unmixing literature, which assumes that, for each pixel, the reflectance spectrum is a linear combination of the spectrum of the endmembers weighted by the corresponding abundances~\cite{bioucas2013hyperspectral,keshava2002spectral}. This model can be described as follows:
\begin{equation}
    \mathbf{Y}=\mathbf{EA}+\mathbf{N}
    \label{eq:Linear model}
\end{equation}
where $\mathbf{Y}=[\mathbf{y}_1,\dots,\mathbf{y}_N]\in \mathbb{R}^{P\times N}$ is a HSI data cube containing the reflectance spectra of $N$ pixels across $P$ spectral bands, i.e., $\mathbf{y}_n\in \mathbb{R}^{P\times 1}$ is the spectra of $n^{th}$ pixel; $\mathbf{E}=[\mathbf{e}_1,...,\mathbf{e}_R]\in \mathbb{R}^{P \times R}$ is the endmember matrix containing the spectral signatures of $R$ endmembers across $P$ spectral bands, i.e., $\mathbf{e}_r\in \mathbb{R}^{P\times 1}$ models the spectra signature of the $r^{th}$ endmember ($r=1,...,R$); $\mathbf{A}=[\mathbf{a}_1,...,\mathbf{a}_N]\in \mathbb{R}^{R \times N}$ is the corresponding fractional abundance matrix, i.e., $\mathbf{a}_n\in \mathbb{R}^{R\times 1}$ is the abundance vector containing the abundances of $R$ different endmembers present in the $n^{th}$ pixel; and $\mathbf{N}\in \mathbb{R}^{P\times N}$ is the additive noise. It is worth noting that in this study, we simplify notation usage by referring to the flattened HSI image $\mathbf{Y}\in \mathbb{R}^{P\times N}$, while working with the HSI image of size $N_1$ by $N_2$, represented by $\mathbf{Y}\in \mathbb{R}^{P\times N_1 \times N_2}$.

Generally, the abundance is subjected to a non-negative constraint (ANC) and a sum-to-one constraint (ASC), i.e., $\mathbf{A}\geq \mathbf{0}$ and $\mathbf{A}^T \mathbf{1}_R=\mathbf{1}_N$, where $\mathbf{1}_R$ is the all one vector with size $R\times 1$. Similarly, the endmember matrix is also subjected to non-negative constraint (ENC), $\mathbf{E}\geq \mathbf{0}$, to be physically meaningful.

The goal of blind linear unmixing is to estimate $\mathbf{E}$ and $\mathbf{A}$ given only $\mathbf{Y}$. A popular approach to address this type of problem involves solving~\cite{Sigurdsson2016}:
\begin{equation}
\begin{aligned}
    \mathbf{\hat{E},\hat{A}}=\textit{arg}\min_{\mathbf{E,A}}{\frac{1}{2}\|\mathbf{Y}-\mathbf{EA}\|_{F}^2+R(\mathbf{A})}\\
    s.t., \mathbf{E}\geq \mathbf{0},\mathbf{A}\geq \mathbf{0}, \mathbf{A}^T \mathbf{1}_R=\mathbf{1}_N
\end{aligned}
\label{eq: BU problem}
\end{equation}
where, $R(\mathbf{A})$ is a regularizer depending on abundance matrix $\mathbf{A}$, such as total variation (TV)~\cite{Sigurdsson2016}. Generally, the choice of $R$ is heavily dependent on prior knowledge about the task at hand. The optimisation problem \eqref{eq: BU problem} is typically solved by the multiplicative update rule~\cite{Feng2018} or a two-stage cyclic descent method~\cite{Sigurdsson2016}, which iteratively fix $\mathbf{A}$, update $\mathbf{E}$, and fix $\mathbf{E}$, update $\mathbf{A}$.

\subsection{Algorithm Unrolling Based Methods}
The algorithm unrolling technique, which is originally proposed in~\cite{gregor2010learning}, solves the sparse coding problem by converting iterative solvers into neural network operations. The first application of this technique in the unmixing problem is the work by~\cite{qian2019deep}. In particular, they first formulated a sparsity-constrained linear regression unmixing problem and solved it using the ISTA solver. Then they unfolded the ISTA solver to design a deep neural network architecture for AE, known as MNN-AE. Later,~\cite{Qian2020} extended it to generate estimations for both abundances and endmembers by appending a linear layer to MNN-AE. The weights of the linear layer are interpreted as the endmembers. This method is known as model inspired neural networks for blind unmixing (MNN-BU). Similarly, as the alternating direction method of multipliers (ADMM) solver delivers fast convergence with satisfactory accuracy~\cite{Tao2015} ,~\cite{Zhou2021} and~\cite{Zhou2021b} tried to unroll the ADMM solvers into neural networks for AE and BU, respectively, which are known as UADMM-AE and UADMM-BU. It is shown that unfolding ADMM-based networks perform better than the ISTA-based ones~\cite{Zhou2021b}. In the meantime, building upon a $L_{p}$ sparsity constrained nonnegative matrix factorization ($L_{p}$-NMF) model,~\cite{Xiong2021} proposed an LMM-based end-to-end deep neural network named SNMF-Net for hyperspectral unmixing. 

Despite the fruitful developments, these unfolding based methods, as other conventional deep learning based unmixing networks, can not guarantee to generate physically meaningful unmixing results~\cite{UnDIP2021}. Another problem of these model-aware learning based approaches is that they are built upon linear models. It is known that convolutional models are very powerful in terms of image processing~\cite{krizhevsky2017imagenet}. It remains unknown how to incorporate convolutional models into the unfolding based unmixing networks.


\subsection{DIP Based Methods}
The deep image prior technique~\cite{DIP2018} is originally proposed to solve image restoration problems and they argue that the explicit regularizers can be dropped as the structure of an image generator can readily capture many image statistics. Later, DIP is introduced to solve linear hyperspectral unmixing problems in~\cite{UnDIP2021}. Specifically, they first rely on the existing geometric endmember extraction method to generate an estimation of endmembers, which is then used to guide the training of a DIP network, known as UnDIP, that estimates the corresponding abundances. This guidance can guarantee the network outputs meaningful abundance estimations. This method, however, has the disadvantage that its performance is limited by the quality of the endmember estimations provided by existing methods~\cite{Zhou2022}. To mitigate this issue,~\cite{Zhou2022} proposed a double DIP structure, known as BUDDIP, that can deliver estimations for both endmembers and abundances. In particular, they propose to use existing unmixing methods to generate both endmember estimations and abundance estimations, which are then used to guide the network training procedure in the form of a composite loss function. It is shown that BUDDIP can not only generate physically meaningful unmixing results but also achieve better unmixing performance than the guidance itself. However, how to design a good DIP architecture for the task at hand remains an open problem. In addition, it has been shown that RED regularizers can improve image restoration quality in single DIP networks~\cite{DeepRED}. Thus, it is also interesting to understand whether the performance of double DIP based unmixing networks can be further improved by adding explicitly RED regularizers.

\section{Proposed Method}
\begin{figure}[!htb]
\centering
\includegraphics{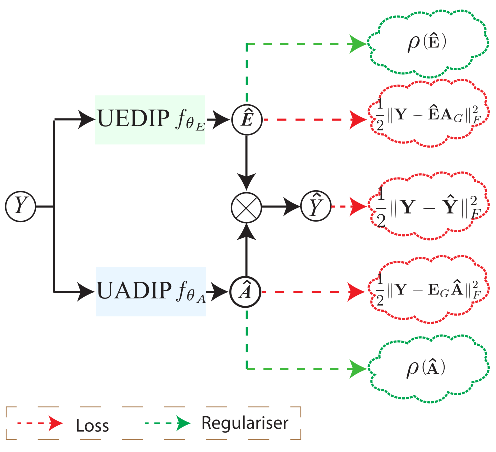}
\caption{NBA network with RED regularizer.}
\label{fig: NBA-net structure}
\end{figure}
We now introduce the proposed network for blind unmixing based on ADMM solvers (NBA) with regularizers by denoising (RED).

The proposed network mainly consists of two modules: UEDIP and UADIP, as illustrated in Fig.~\ref{fig: NBA-net structure}. UEDIP and UADIP both take the observations $\mathbf{Y}$ as input and deliver the endmember estimation $\hat{\mathbf{E}}$ and abundance estimation $\hat{\mathbf{A}}$, respectively. Based on the LMM~\eqref{eq:Linear model}, the network can immediately generate an reconstruction of observations $\hat{\mathbf{Y}}$. In order to generate physically meaningful outputs, we propose to train the network using a composite loss function. Additionally, we also find that the performance can be further improved by adding explicit regularizers RED to both $\hat{\mathbf{E}}$ and $\hat{\mathbf{A}}$.

Before we introduce how to construct UEDIP and UADIP, we first introduce the proposed MCU approach for EE and AE, which is the stepping stone to our network constructions.
\subsection{MatrixConv Unmixing (MCU) Approach}
\label{sec: MCU-based HSI model}
We now introduce the proposed MCU approach for EE and AE task.
\subsubsection{MCU-based EE}
The blind unmixing problem reduces to the EE problem when the abundance $\mathbf{A}$ is known as a priori information, which can be either given as ground truth or estimated by existing methods. The target of the EE problem amounts to solving
\begin{equation}
\label{eq: Endmember estimation problem}
    \min_{\mathbf{E}} {\frac{1}{2}\|\mathbf{Y}^T-\mathbf{A}^T\mathbf{E}^T\|_F^2},\quad s.t.,\mathbf{E}\geq \mathbf{0}
\end{equation}
where we have taken the transpose form of \eqref{eq: BU problem} to simplify the derivation below. Suppose that $\mathbf{E}^T$ is the result of convolutional sparse coding as follows:
\begin{equation}
    \label{eq: Endm CSC model}
    \mathbf{E}^T=\sum_{i=1}^{m_{E}} \mathbf{d}_{i}^{E}\ast \mathbf{\alpha}_{i}^{E}
\end{equation}
where, $\mathbf{d}_{i}^{E}$ and $\mathbf{\alpha}_{i}^{E}$ are the $i^{th}$ 1D convolutional kernel of size $k_E$ and the corresponding sparse code, respectively. $m_E$ is the number of convolutional kernels. As convolution is a linear operation, it can be converted into a linear operator as follows:
\begin{equation}
\label{eq: conv to linear}
    \sum_{i=1}^{m_{E}} \mathbf{d}_{i}^{E}\ast \mathbf{\alpha}_{i}^{E}=\mathbf{D}_{E}\mathbf{\Gamma}_{E}
\end{equation}
where, $\mathbf{D}_{E}$ and $\mathbf{\Gamma}_{E}$ are the convolutional dictionary and corresponding sparse code that encompass $\{\mathbf{d}_{i}^{E}\}_{i=1}^{m_{E}}$ and $\{\mathbf{\alpha}_{i}^{E}\}_{i=1}^{m_{E}}$. Suppose $\mathbf{D}_{E}$ is known, then \eqref{eq: Endmember estimation problem} becomes an MCU problem:
\begin{equation}
\label{eq: convolutional Endmember sparse code problem}
    \min_{\mathbf{\Gamma}_{E}}{\frac{1}{2}\|\mathbf{Y}^T-\mathbf{A}^T\mathbf{D}_{E}\mathbf{\Gamma}_{E}\|_F^2}+\lambda\|\mathbf{\Gamma}_{E}\|_{1}
\end{equation}
which can be solved via an iterative update algorithm such as ISTA or ADMM. In this work, we use the ADMM solver as it has been shown unfolding ADMM based network achieve better performance than unfolding ISTA based network~\cite{Zhou2021b}. The augmented lagrangian form of \eqref{eq: convolutional Endmember sparse code problem} is:
\begin{equation}
\begin{aligned}
    \min_{\mathbf{\Gamma}_{E},\mathbf{\Omega}_{E}}&{\frac{1}{2}\|\mathbf{Y}^T-\mathbf{A}^T\mathbf{D}_{E}\mathbf{\Omega}_{E}\|_F^2+\lambda\|\mathbf{\Gamma}_{E}\|_{1}}\\
    &+\frac{\rho_E}{2}\|\mathbf{\Gamma}_{E}-\mathbf{\Omega}_{E}+\mathbf{u}_{E}\|_F^2
\end{aligned}
\end{equation}
where $\mathbf{\Omega}_{E},\mathbf{u}_{E},\rho_E$ are variables introduced by the ADMM algorithm. The ADMM solver for the problem above is as follows:
\begin{equation}
\label{eq: ADMM for MCU-EE model}
    \begin{cases}
    \begin{aligned}
\mathbf{\Omega}_{E}^{j+1}=&\big((\mathbf{A}^T\mathbf{D}_{E}\big)^T (\mathbf{A}^T\mathbf{D}_{E})+\rho_E\mathbf{I})^{-1}\\
&\big((\mathbf{A}^T\mathbf{D}_{E})^T \mathbf{Y}^T+\rho_E(\mathbf{\Gamma}_{E}^{j}+\mathbf{u}_{E}^{j} )\big)\end{aligned}\\
    \mathbf{\Gamma}_{E}^{j+1}=\textit{Soft}_{\frac{\lambda}{L}}\big((1-\frac{\rho_E}{L})\mathbf{\Gamma}_{E}^{j}+\frac{\rho_E}{L}(\mathbf{\Omega}_{E}^{j+1}-\mathbf{u}_{E}^{j})\big)\\
    \mathbf{u}_{E}^{j+1}=\mathbf{u}_{E}^{j}+(\mathbf{\Gamma}_{E}^{j+1}-\mathbf{\Omega}_{E}^{j+1})
    \end{cases}
\end{equation}
where $\textit{Soft}_{z}(x)=sign(x)\cdot(|x|-z)_{+}$ is the element-wise soft-threshold operator.

\subsubsection{MCU-based AE}
Similarly, when the endmember $\mathbf{E}$ is known as a priori information, the blind unmixing problem reduces to the AE problem, which aims at solving
\begin{equation}
\label{eq: Abundance estimation problem}
    \min_{\mathbf{A}} {\frac{1}{2}\|\mathbf{Y}-\mathbf{E}\mathbf{A}\|_F^2},\quad s.t.,\mathbf{A}\geq \mathbf{0}, \mathbf{A}^T \mathbf{1}_r=\mathbf{1}_n
\end{equation} 

Note that the regularization term in~\eqref{eq: BU problem} can be dropped according to DIP techniques~\cite{DIP2018}. Slightly different from the case of unfolding based EDIP, here we assume the abundance is the result of nonnegative convolutional sparse coding as follows:
\begin{equation}
    \label{eq: Abundance NCSC model}
    \mathbf{A}=\sum_{i=1}^{m_{A}} \mathbf{d}_{i}^{A}\ast \mathbf{\alpha}_{i}^{A}, \text{ where } \mathbf{\alpha}_{i}^{A}\geq 0,\; \forall i 
\end{equation}
where, $\mathbf{d}_{i}^{A}$ and $\mathbf{\alpha}_{i}^{A}$ are the $i^{th}$ 2D convolutional kernel of size $k_A\times k_A$ and the corresponding sparse code, respectively. $m_A$ is the number of convolutional kernels. We adopt the nonnegative sparse code here because we experimentally find that it achieves better performance in the abundance estimation problem, which could be due to the fact that the abundance naturally has the ANC constraint. Again, since convolution is a linear operation, we can convert it into the form as follows:
\begin{equation}
    \sum_{i=1}^{m_{A}} \mathbf{d}_{i}^{A}\ast \mathbf{\alpha}_{i}^{A}=\mathbf{D}_{A}\mathbf{\Gamma}_{A}
\end{equation}
where $\mathbf{D}_A$ is a matrix that encodes the set of convolutional kernels $\{\mathbf{d}_{i}^{A}\}$ and $\mathbf{\Gamma}_{A}$ is the corresponding sparse code that encodes $\{\mathbf{\alpha}_{i}^{A}\}$. Suppose $\mathbf{D}_A$ is known, then~\eqref{eq: Abundance estimation problem} becomes an MCU problem:
\begin{equation}
\label{eq: convolutional Abundance sparse code problem}
    \min_{\mathbf{\Gamma}_{A}}{\frac{1}{2}\|\mathbf{Y}-\mathbf{E}\mathbf{D}_{A}\mathbf{\Gamma}_{A}\|_F^2}+\lambda\|\mathbf{\Gamma}_{A}\|_{1},\quad s.t., \mathbf{\Gamma}_{A}\geq \mathbf{0}
\end{equation}
Similarly, by introducing the auxiliary variables $\mathbf{\Omega}_{A},\mathbf{u}_{A},\rho_A$, the augmented Lagrangian of~\eqref{eq: convolutional Abundance sparse code problem} is
\begin{equation}
\begin{aligned}
    \min_{\mathbf{\Gamma}_{A},\mathbf{\Omega}_{A}}&{\frac{1}{2}\|\mathbf{Y}-\mathbf{E}\mathbf{D}_{A}\mathbf{\Omega}_{A}\|_F^2+\lambda\|\mathbf{\Gamma}_{A}\|_{1}}\\
    &+R_{+}(\mathbf{\Gamma}_{A})+\frac{\rho_A}{2}\|\mathbf{\Gamma}_{A}-\mathbf{\Omega}_{A}+\mathbf{u}_{A}\|_F^2
\end{aligned}
\label{eq: AL of convolutional Abundance sparse code problem}
\end{equation}
where $R_{+}(\mathbf{\Gamma}_{A})$ is the non-negative constraint. This problem can be solved via the ADMM solver as follows:
\begin{equation}
\label{eq: ADMM for MCU-AE model}
    \begin{cases}
    \begin{aligned}
\mathbf{\Omega}_{A}^{j+1}=&\big((\mathbf{E}\mathbf{D}_{A}\big)^T (\mathbf{E}\mathbf{D}_{A})+\rho_A\mathbf{I})^{-1}\\
&\big((\mathbf{E}\mathbf{D}_{A})^T \mathbf{Y}+\rho_A(\mathbf{\Gamma}_{A}^{j}+\mathbf{u}_{A}^{j} )\big)\end{aligned}\\
    \mathbf{\Gamma}_{A}^{j+1}=\max{\Big(\textit{Soft}_{\frac{\lambda}{L}}\big((1-\frac{\rho_A}{L})\mathbf{\Gamma}_{A}^{j}+\frac{\rho_A}{L}(\mathbf{\Omega}_{A}^{j+1}-\mathbf{u}_{A}^{j})\big),0\Big)}\\
    \mathbf{u}_{A}^{j+1}=\mathbf{u}_{A}^{j}+(\mathbf{\Gamma}_{A}^{j+1}-\mathbf{\Omega}_{A}^{j+1})
    \end{cases}
\end{equation}

\subsection{Unfolding based EDIP}
\label{sec: Unfolding based EDIP}
We now introduce how to construct UEDIP network structure by unfolding the ADMM solver~\eqref{eq: ADMM for MCU-EE model}.

\begin{figure}[!htb]
\centering
\includegraphics{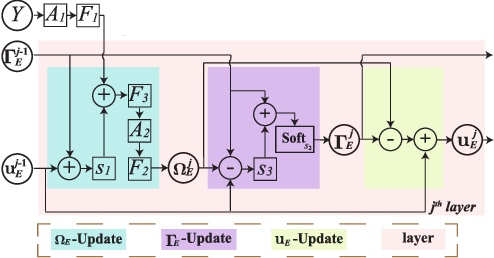}
\caption{UEDIP layer-wise operations.}
\label{fig: UEDIP layer structure}
\end{figure}

According to unfolding techniques, let's define the learnable parameters $\mathbf{A}_1,\mathbf{A}_2,\mathbf{F}_1,\mathbf{F}_2,\mathbf{F}_3, s_1,s_2,s_3$, such that, $\mathbf{F}_1\ast\mathbf{A}_1$ and $\mathbf{F}_2\ast\mathbf{A}_2\times\mathbf{F}_3$ can replace the role of $(\mathbf{A}^T\mathbf{D}_{E})^T$ and $\big((\mathbf{A}^T\mathbf{D}_{E}\big)^T (\mathbf{A}^T\mathbf{D}_{E})+\rho_E\mathbf{I})^{-1}$ in~\eqref{eq: ADMM for MCU-EE model}, and $s_1,s_2,s_3$ can play the role of $\rho_E,\lambda/L,\rho_E/L$ in~\eqref{eq: ADMM for MCU-EE model}, respectively. Note that the parameters $\mathbf{A}_1,\mathbf{A}_2$ are the linear matrix, while $\mathbf{F}_1,\mathbf{F}_2,\mathbf{F}_3$ correspond to convolutional dictionaries. Then a layer of network operation, which is shown in Fig.~\ref{fig: UEDIP layer structure}, is defined as follows:
\begin{equation}
    \begin{cases}
    \mathbf{\Omega}_{E}^{j+1}=\mathbf{F}_2\ast\mathbf{A}_2\times\mathbf{F}_3\ast\big(\mathbf{F}_1\ast\mathbf{A}_1\times \mathbf{Y}^T+s_1(\mathbf{\Gamma}_{E}^{j}+\mathbf{u}_{E}^{j} )\big)
\\
    \mathbf{\Gamma}_{E}^{j+1}=\textit{Soft}_{s_2}\big((1-s_3)\mathbf{\Gamma}_{E}^{j}+s_3(\mathbf{\Omega}_{E}^{j+1}-\mathbf{u}_{E}^{j})\big)\\
    \mathbf{u}_{E}^{j+1}=\mathbf{u}_{E}^{j}+(\mathbf{\Gamma}_{E}^{j+1}-\mathbf{\Omega}_{E}^{j+1})
    \end{cases}
\end{equation}

 where, $\mathbf{\Omega}_{E}^{j}$ is the value of $\mathbf{\Omega}_{E}$ at $j$-th layer, and $j=1,\dots,J_{E}$. Note that we exclude $\mathbf{F}_1\ast\mathbf{A}_1\times \mathbf{Y}^T$ out of the layer-wise operations as we share this value for different layers to reduce the number of learnable parameters. It can be seen that each layer would output updated estimations of $\mathbf{\Gamma}_{E}$ and $\mathbf{u}_{E}$. It is worth mentioning that this structure is different from the structure of a general-purpose DNN or CNN, as it interweaves matrix multiplication, such as $\mathbf{A}_1,\mathbf{A}_2$ with convolutions $\mathbf{F}_1,\mathbf{F}_2,\mathbf{F}_3$. And in a general CNN, the linear matrix multiplication is usually appended after a stack of convolution layers, serving as output layers.

\begin{figure}[!htb]
\centering
\includegraphics{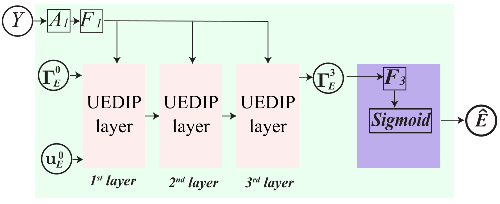}
\caption{An illustration of 3-layer UEDIP structure. It takes $\mathbf{Y}$ as input and outputs endmember estimation $\hat{\mathbf{E}}$. The operation in each layer is shown in Fig.~\ref{fig: UEDIP layer structure}.}
\label{fig: unfolding EDIP}
\end{figure}

An illustration of $J_E=3$-layer unfolding ADMM based EDIP (UEDIP) structure is shown in Fig.~\ref{fig: unfolding EDIP}. It takes $\mathbf{Y}$ as input and outputs endmember estimation $\hat{\mathbf{E}}$. Note that for the first layer, we use the all zero initialisation for $\mathbf{\Gamma}_{E}^{0}$ and $\mathbf{u}_{E}^{0}$. In the last layer, after generating the sparse code $\mathbf{\Gamma}_{E}^{3}$, according to our approach~\eqref{eq: Endm CSC model}, the endmember is estimated by further going through a convolutional operator $\mathbf{F}_3$ as shown in Fig.~\ref{fig: unfolding EDIP}. Finally, a Sigmoid activation is added to impose the ENC constraint. The learnable parameters in each layer are unshared. The UEDIP network is denoted as $f_{\boldsymbol{\theta}_E}(\mathbf{Y})$, where $\boldsymbol{\theta}_E=\{\mathbf{A}_1,\mathbf{A}_2,\mathbf{F}_1,\mathbf{F}_2,\mathbf{F}_3, s_1,s_2,s_3\}$ is the learnable parameters in UEDIP.


\subsection{Unfolding based ADIP}
\label{sec: Unfolding based ADIP}
We now introduce how to construct UADIP structure via unfolding the ADMM solver~\eqref{eq: ADMM for MCU-AE model}.

According to unfolding techniques, let's define the learnable parameters $\mathbf{E}_1,\mathbf{E}_2,\mathbf{U}_1,\mathbf{U}_2,\mathbf{U}_3, v_1,v_2,v_3$, such that, $\mathbf{U}_1\ast\mathbf{E}_1$ and $\mathbf{U}_2\ast\mathbf{E}_2\times\mathbf{U}_3$ can replace the role of $(\mathbf{E}\mathbf{D}_{A})^T$ and $\big((\mathbf{E}\mathbf{D}_{A}\big)^T (\mathbf{E}\mathbf{D}_{A})+\rho_A\mathbf{I})^{-1}$ in~\eqref{eq: ADMM for MCU-AE model}, and $v_1,v_2,v_3$ can play the role of $\rho_A,\lambda/L,\rho_A/L$ in~\eqref{eq: ADMM for MCU-AE model}, respectively. Note that $\max{(\textit{Soft}(\cdot))}$ is equivalent to a shifted $\textit{ReLU}$, then a layer of network operation is defined as follows:
\begin{equation}
    \begin{cases}
    \mathbf{\Omega}_{A}^{j+1}=\mathbf{U}_2\ast\mathbf{E}_2\times\mathbf{U}_3\ast\big(\mathbf{U}_1\ast\mathbf{E}_1\times \mathbf{Y}+v_1(\mathbf{\Gamma}_{A}^{j}+\mathbf{u}_{A}^{j} )\big)
\\
    \mathbf{\Gamma}_{A}^{j+1}=\textit{ReLU}\big((1-v_3)\mathbf{\Gamma}_{A}^{j}+v_3(\mathbf{\Omega}_{A}^{j+1}-\mathbf{u}_{A}^{j})-v_2\big)\\
    \mathbf{u}_{A}^{j+1}=\mathbf{u}_{A}^{j}+(\mathbf{\Gamma}_{A}^{j+1}-\mathbf{\Omega}_{A}^{j+1})
    \end{cases}
\end{equation}

\begin{figure}[!htb]
\centering
\includegraphics{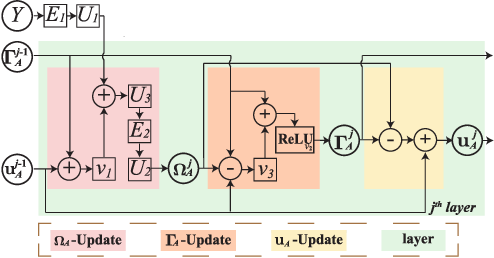}
\caption{UADIP layer-wise operations.}
\label{fig: UADIP layer structure}
\end{figure}

This layer-wise operation is shown in Fig.~\ref{fig: UADIP layer structure}. Similarly, we exclude out $\mathbf{U}_1\ast\mathbf{E}_1\times \mathbf{Y}$, which is shared in different layers. Each layer would generate updated estimations of $\mathbf{\Gamma}_{A}$ and $\mathbf{u}_{A}$.

\begin{figure}[!htb]
\centering
\includegraphics{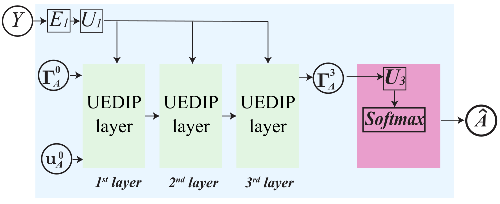}
\caption{An illustration of 3-layer UADIP structure. It takes $\mathbf{Y}$ as input and outputs abundance estimation $\hat{\mathbf{A}}$. The operation in each layer is shown in Fig.~\ref{fig: UADIP layer structure}.}
\label{fig: unfolding ADIP}
\end{figure}

An illustration of $J_A=3$-layer unfolding ADMM based ADIP (UADIP) structure is shown in Fig.~\ref{fig: unfolding ADIP}. It takes $\mathbf{Y}$ as input and outputs abundance estimation $\hat{\mathbf{A}}$. Again, in the first layer, we use all zero initialisation for $\mathbf{\Gamma}_{A}^{0}$ and $\mathbf{u}_{A}^{0}$. In the last layer, after generating the sparse code $\mathbf{\Gamma}_{A}^{3}$, according to our approach~\eqref{eq: Abundance NCSC model}, the abundance is estimated by further going through a convolutional operator $\mathbf{U}_3$. Finally, a Softmax activation is added to impose the ASC and ANC constraints. Akin to UEDIP, we use the unshared parameterisation strategy for different layers in UADIP. The UADIP network is denoted as $f_{\boldsymbol{\theta}_A}(\mathbf{Y})$, where $\boldsymbol{\theta}_A$ is the learnable parameters in UADIP.


\subsection{Network for Blind-unmixing Based on ADMM (NBA)}
In this section, we combine the proposed UEDIP and UADIP, giving rise to the double DIP Network for Blind unmixing based on ADMM solver (NBA). Specifically, according to the model~\eqref{eq:Linear model}, after gaining an estimation of endmember and abundance, $\hat{\mathbf{E}}$ and $\hat{\mathbf{A}}$, using UEDIP and UADIP, respectively, we can immediately generate a reconstruction of the HSI observations, as follows:
\begin{equation}
    \hat{\mathbf{Y}}=\hat{\mathbf{E}}\hat{\mathbf{A}}
\end{equation}
Hence, the overall structure of the proposed NBA network is obtained by assembling UEDIP and UADIP together, as shown in Fig.~\ref{fig: NBA-net structure}. Specifically, we denote all the learnable parameters in NBA network as $\mathbf{\Theta}=\{\boldsymbol{\theta}_E,\boldsymbol{\theta}_A\}$, where $\boldsymbol{\theta}_E,\boldsymbol{\theta}_A$ are the learnable parameters in UEDIP and UADIP, respectively.

\textit{Training details.} According to the discussion above, given the HSI image $\mathbf{Y}$, the NBA network delivers three outputs: endmember estimation $\hat{\mathbf{E}}$, abundance estimation $\hat{\mathbf{A}}$ and HSI reconstructions $\hat{\mathbf{Y}}$. Correspondingly, we propose to train the NBA network, in an end-to-end manner, using three loss terms. 

The first loss term is imposed on endmember estimation by UEDIP, $\hat{\mathbf{E}}=f_{\boldsymbol{\theta}_E}(\mathbf{Y})$. We propose to use the following loss function:
\begin{equation}
    L_{E}(\boldsymbol{\theta}_E)=\frac{1}{2}\|\mathbf{Y}-f_{\boldsymbol{\theta}_E}(\mathbf{Y})\mathbf{A}_{G}\|_F^2
\end{equation}
where, $\mathbf{A}_{G}$ is the training guidance generated by some existing unmixing methods such as FCLS~\cite{Heinz2001}.This loss term would guarantee UEDIP network $f_{\boldsymbol{\theta}_E}$ to generate meaningful estimation of endmembers. 

Similarly, for abundance estimation by UADIP, $\hat{\mathbf{A}}=f_{\boldsymbol{\theta}_A}(\mathbf{Y})$, we propose the loss term, given by 
\begin{equation}
    L_{A}(\boldsymbol{\theta}_A)=\frac{1}{2}\|\mathbf{Y}-\mathbf{E}_{G}f_{\boldsymbol{\theta}_A}(\mathbf{Y})\|_F^2
\end{equation}
where, $\mathbf{E}_{G}$ is an estimation of endmembers generated by some existing unmixing methods such as SiVM~\cite{Heylen2011}, which would guide the network training to generate meaningful estimation of abundances. 

Finally, for the reconstruction of HSI observations $\hat{\mathbf{Y}}$, we propose the third loss term given by 
\begin{equation}
    L_{BU}(\mathbf{\Theta})=\frac{1}{2}\|\mathbf{Y}-\hat{\mathbf{Y}}\|_F^2
\end{equation}
This term enforces the NBA network to search for better unmixing results than the guidance $\mathbf{E}_{G},\mathbf{A}_{G}$. Without this term, the network would simply yield endmember and abundance estimates close to the guidance $\mathbf{E}_{G},\mathbf{A}_{G}$.

By assembling the loss terms above, we present the composite loss functions given by
\begin{equation}
    \label{eq: NBA loss}
    L(\mathbf{\Theta})=\alpha_1\cdot L_{E}+\alpha_2\cdot L_{A}+\alpha_3\cdot L_{BU}
\end{equation}
where $\alpha_1,\alpha_2,\alpha_3$ are loss weights that control the relative importance of corresponding loss terms.


\subsection{Explicit regularization}
\label{sec: regularizer}
\begin{algorithm}
\caption{ADMM solver of NBA with RED regularizers}\label{alg:alg1}
\begin{algorithmic}[1]
\State \textbf{Parameters}:
\vspace{-0.5em}
\begin{itemize}
    \setlength\itemsep{0em}
    \item $\alpha_{1}\sim\alpha_5$ - loss weights
    \item $\mu_E,\mu_A$ - ADMM hyper-parameters
    \item $T$ - the maximum number of ADMM iterations.
\end{itemize}
\vspace{-0.5em}
\State \textbf{Init:} $\mathbf{d}_{E}^{0},\mathbf{d}_{A}^{0},\mathbf{X}_{E}^{0},\mathbf{X}_{A}^{0}\leftarrow 0,0,0,0$, $t\leftarrow 0$, and initialize NBA network randomly.
   \Repeat
   \State \textbf{Update $\mathbf{\Theta}$: } solving~\eqref{eq: ADMM eq1} via training NBA network.
   \State \textbf{Update $\mathbf{X}_{E},\mathbf{X}_{A}$: } using fixed point strategy~\eqref{eq: fixed-point update}.
   \State \textbf{Update $\mathbf{d}_{E},\mathbf{d}_{A}$: } using~\eqref{eq: ADMM eq3}.
   \State $t\leftarrow t+1$
   \Until{converged or $t>T$}
\end{algorithmic}
\end{algorithm}

In~\cite{DIP2018}, the DIP technique suggests that the explicit regularizer $R(\cdot)$ in a recovery problem~\eqref{eq: BU problem} can be dropped and the neural network parameterisation itself can implicitly capture the prior. However,~\cite{DeepRED} found that the performance can be further boosted by adding an explicit RED regularizer~\cite{RED} to the DIP network. This is because RED can offer an extra explicit regularization in addition to the implicit one. In this work, we move one step further by introducing two RED regularizers into the double DIP network structures to further improve network performance. Specifically, we propose adding explicit regularization RED into~\eqref{eq: NBA loss} for both $\hat{\mathbf{E}}$ and $\hat{\mathbf{A}}$, then the loss function becomes:
\begin{equation}
    \label{eq: BUDIPRED loss}
    \mathcal{L_{R}}=L+\alpha_4\cdot\rho(\hat{\mathbf{E}})+\alpha_5\cdot\rho(\hat{\mathbf{A}})
\end{equation}
where, we have dropped the parameters $\mathbf{\Theta}$ in $L$ for notation simplicity, and $\rho(\mathbf{X})=\frac{1}{2}\mathbf{X}^T(\mathbf{X}-f_D(\mathbf{X}))$ is the RED regularizer with denoiser function $f_D(\cdot)$, which could be any well-developed denoiser such as Non-Local Means (NLM)~\cite{NLM}, BM3D~\cite{BM3D} and so on. It can be seen that RED tries to minimize the inner product between the image $\mathbf{X}$ and its denoising residual. It is essentially derived from an image-adaptive Laplacian, which in turn depends on the choice of denoiser $f_D$. Thus, RED can represent a variety of regularizations by incorporating different image denoisers. In this work, we use the NLM denoiser as our default settings. It should be noted that the RED regularizer is proposed to deal with image data which is embedded in 3D space. The endmember however is embedded in 2D space, $\mathbf{E}\in \mathbb{R}^{P\times R}$. In this work, we provide a simple workaround by expanding it to 3D space $\mathbb{R}^{1\times P\times R}$ and treat it as a gray-image like data. 

After adding the RED regularizers, it is not simple to directly train the network with objective~\eqref{eq: BUDIPRED loss} as it will require many iterations to converge~\cite{RED}. In this work, we propose to solve~\eqref{eq: BUDIPRED loss} using the ADMM solver~\cite{RED,DeepRED}. We first re-write~\eqref{eq: BUDIPRED loss} in the scaled form of the Augmented Lagrangian (AL):
\begin{equation}
    \label{eq: scaled AL for NBARED}
    \begin{aligned}
    L+\alpha_4\cdot\rho(\mathbf{X}_{E})+\frac{\mu_{E}}{2}\|\mathbf{X}_{E}-\hat{\mathbf{E}}-\mathbf{d}_{E}\|_F^2\\
    +\alpha_5\cdot\rho(\mathbf{X}_{A})+\frac{\mu_{A}}{2}\|\mathbf{X}_{A}-\hat{\mathbf{A}}-\mathbf{d}_{A}\|_F^2
    \end{aligned}
\end{equation}
where, $\{\mathbf{X}_{E},\mathbf{X}_{A},\mathbf{d}_{E},\mathbf{d}_{A}\}$ are auxiliary variables introduced via AL, and $\mu_{E},\mu_{A}$ are also the hyperparameters of AL. The ADMM solver to~\eqref{eq: scaled AL for NBARED} gives the following iterative update rules:
\begin{numcases}{}
   \begin{aligned}
    \mathbf{\Theta}^{t+1}=\textit{arg}\min_{\mathbf{\Theta}}{L+\frac{\mu_{E}}{2}\|\mathbf{X}_{E}^{t}-\hat{\mathbf{E}}-\mathbf{d}_{E}^{t}\|_F^2}\\
    +\frac{\mu_{A}}{2}\|\mathbf{X}_{A}^{t}-\hat{\mathbf{A}}-\mathbf{d}_{A}^{t}\|_F^2
    \end{aligned} \label{eq: ADMM eq1}
   \\
   \begin{aligned}
    \mathbf{X}_{E}^{t+1},\mathbf{X}_{A}^{t+1}=\textit{arg}\min_{\mathbf{X}_{E},\mathbf{X}_{A}}{\alpha_4\cdot\rho(\mathbf{X}_{E})+\alpha_5\cdot\rho(\mathbf{X}_{A})}\\
    +\frac{\mu_{E}}{2}\|\mathbf{X}_{E}-\hat{\mathbf{E}}^{t+1}-\mathbf{d}_{E}^{t}\|_F^2\\
    +\frac{\mu_{A}}{2}\|\mathbf{X}_{A}^{t}-\hat{\mathbf{A}}^{t+1}-\mathbf{d}_{A}^{t}\|_F^2
    \end{aligned} \label{eq: ADMM eq2}
    \\
    \begin{aligned}
    \{\mathbf{d}_{E}^{t+1},\mathbf{d}_{A}^{t+1}\}=\{&\mathbf{d}_{E}^{t}+(\hat{\mathbf{E}}^{t+1}-\mathbf{X}_{E}^{t+1}),\\
    &\mathbf{d}_{A}^{t}+(\hat{\mathbf{A}}^{t+1}-\mathbf{X}_{A}^{t+1})\}
    \end{aligned} \label{eq: ADMM eq3}
\end{numcases}
 The first objective~\eqref{eq: ADMM eq1} is optimized through network training. Note that $\hat{\mathbf{E}},\hat{\mathbf{A}}$ are the outputs of the network which also depend on $\mathbf{\Theta}$. The second objective~\eqref{eq: ADMM eq2} can be solved via either a fixed-point based solution or gradient-descent based solution~\cite{RED,DeepRED}. In this work, we use the fixed-point based solution given by
\begin{equation}
\label{eq: fixed-point update}
    \begin{cases}
    \mathbf{X}_{E}^{t+1}=\frac{1}{\alpha_4+\mu_{E}}(\alpha_4f_D(\mathbf{X}_{E}^{t})+\mu_{E}(\hat{\mathbf{E}}^{t+1}+\mathbf{d}_{E}^{t}))\\
    \mathbf{X}_{A}^{t+1}=\frac{1}{\alpha_5+\mu_{A}}(\alpha_5f_D(\mathbf{X}_{A}^{t})+\mu_{A}(\hat{\mathbf{A}}^{t+1}+\mathbf{d}_{A}^{t}))
    \end{cases}
\end{equation}
We summarize the overall steps of the proposed approach in Algorithm~\ref{alg:alg1}.


\section{Experiments}
In this section, we show the effectiveness of the proposed method by evaluating on both synthetic and real datasets. 
\subsection{Metrics}
To fully show the effectiveness, we use different metrics in the literature to evaluate the performance. Specifically, to measure the quality of abundance estimation between the ground truth abundance vector $\mathbf{a}_{n}\in \mathbb{R}^{R\times 1}$ for $n^{th}$ pixel and the corresponding estimation $\hat{\mathbf{a}}_{n}$, we adopt the well-known root mean square error (RMSE) and abundance angle distance (AAD)~\cite{Miao2007}. These metrics are measured as follows:
\begin{align}
    \label{metrics: abundance}
    RMSE_n&=\sqrt{\frac{1}{R}\sum_{r =1}^{R}(a_{r,n}-\hat{a}_{r,n})^{2}}\\
    AAD_n&=\frac{180}{\pi}\cos^{-1}{\left(\frac{\mathbf{a}_{n}^{T}\hat{\mathbf{a}}_n}{\|\mathbf{a}_{n}\|_{2}\|\hat{\mathbf{a}}_n\|_{2}}\right)}
\end{align}
The final scalar quantities are obtained by averaging over all pixels.

Regarding the endmember estimation, we follow the popular spectral angle distance (SAD) to measure the quality of an estimated endmember signature $\hat{\mathbf{e}}_r$ from its ground truth $\mathbf{e}_r$, which is given by:
\begin{equation}
\label{metrics: endmember}
    SAD_r=\frac{180}{\pi}\cos^{-1}{\left(\frac{\mathbf{e}_{r}^{T}\hat{\mathbf{e}}_r}{\|\mathbf{e}_{r}\|_{2}\|\hat{\mathbf{e}}_r\|_{2}}\right)}
\end{equation}
This metric is further averaged over different endmembers to yield the final scalar quantity.

\subsection{Datasets}
We adopt both synthetic and real datasets in this work to compare the performance of the proposed method with other state-of-the-art approaches.
\subsubsection{Synthetic Dataset}
\begin{figure}[!htb]
\centering
\includegraphics{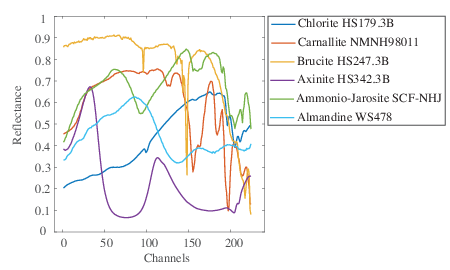}
\caption{Endmember signatures for synthetic data.}
\label{fig:6 endm signatures}
\end{figure}

In this work, we choose the same HSI data synthesis procedure as~\cite{Zhou2021} to generate the synthetic HSI dataset. The procedure involves the following steps:
\begin{itemize}
    \item \textit{Endmember generation.} We randomly select mineral signatures from the famous USGS spectral library to generate the endmember signature matrix. The library, known as splib06~\cite{USGS}, collects spectral reflectances of various minerals measured over 224 channels, ranging from 0.4 $\mu$m to 2.5 $\mu$m. Six spectral signatures are randomly selected from the library, shown in Fig.~\ref{fig:6 endm signatures}, to form a $224 \times 6$ endmember matrix.
    \item \textit{Abundance generation.}
    The abundances underlying each pixel of the HSI image are generated as follows. First, $a^2$ disjoint patches of size $a \times a$ pixels are obtained by dividing a synthetic abundance map of size $a^2 \times a^2$ pixels. Second, for each patch, we randomly select two candidate endmembers out of the six spectral signatures, which are allocated with fractions $\gamma$ and $1-\gamma$, while the remaining four endmembers are assigned with value $0$. Third, the abundance map is convolved with a Gaussian filter of size $(a+1)\times (a+1)$, with variance set to be 2. Finally, a pixel-wise re-scaling is applied to the abundance map to meet the ASC constraint. In this paper, we set $a=10$ and $\gamma=0.8$.
    \item \textit{Noise contamination.} 
    Finally, the additive white gaussian noise (AWGN) is applied to the generated HSI data, leading to the signal-to-noise ratio (SNR), defined as $SNR=10 \log_{10}{\left(E[\mathbf{x}^T\mathbf{x}]/E[\mathbf{n}^T\mathbf{n}]\right)}$, where $\mathbf{x}$ is the clean HSI data, and $\mathbf{n}$ is the noise.
\end{itemize}

\subsubsection{Real Dataset}
\begin{figure}[!htb]
\centering
\subfloat[Jasper Ridge]{ \includegraphics[scale=0.7]{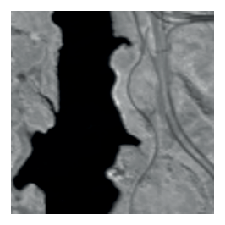}
\label{fig:Image of JasperRidge at 80th channel}}
\hfil
\subfloat[Samson]{\includegraphics{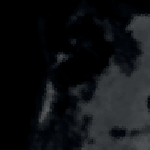}
\label{fig:Image of Samson at 80th channel}}
\caption{HSI image at 80th channel. (a) Jasper Ridge. (b) Samson.}
\label{fig:Image of JasperRridge and Samson at 80th channel.}
\end{figure}
We also evaluate the performance of various approaches using two popular real HSI datasets.
\begin{itemize}
    \item \textit{Jasper Ridge},~\cite{Zhu2014b}, which consists of an HSI image of size $512 \times 614$ pixels. This dataset contains spectral information measured in 224 channels, ranging from 380 nm up to 2500 nm, with a spectral resolution of 9.46 nm. This dataset contains four endmembers: Road, Soil, Water, and Tree. Generally, in order to reduce computational complexity, a sub-image of $100 \times 100$ pixels is considered to deploy faster experimental studies. After removing channels that are affected by dense water vapour and atmospheric effects: channels 1-3, 108-112, 154-166 and 220-224, we only consider 198 out of the 224 channels. A representative image -- associated with the 80th channel image -- is shown in Fig.~\ref{fig:Image of JasperRidge at 80th channel}.
    
    \item \textit{Samson}~\cite{feng2022hyperspectral}, a real dataset characterized by an image resolution of 952×952 pixels. Each pixel is captured at 156 different channels, spanning wavelengths from 401 nm to 889 nm. Notably, the dataset has a high spectral resolution of 3.13 nm. Due to the computational expense associated with the original large image, it is common that a specific region measuring 95×95 pixels is considered. This region commences at the (252,332) pixel of the original image. The image encompasses three distinct endmembers: Soil, Tree, and Water. Fig.~\ref{fig:Image of Samson at 80th channel} visually presents the Samson dataset at the 80th channel.
\end{itemize}
\subsection{Ablation Study}
\label{sec: ablation study}
We first provide ablation studies conducted on the synthetic dataset. Precisely, the synthetic HSI image consists of $100\times100$ pixels, which are further contaminated with AWGN leading to $SNR=30$ dB. By default, the UADIP has $J_A=3$ layers and $m_A=128$ kernels with the kernel size $k_A=5$. Similarly, the UEDIP has $J_E=1$ layer and $m_E=128$ kernels with the kernel size $k_E=5$. We use the well-known Adam optimizer to train our networks with a learning rate set to $1e-3$, and the number of epochs set to 5000. We set the hyperparameter $\beta_2$ of the ADAM optimizer to $0.85$ because we experimentally find that it gives better performance. By default, the NBA network is trained with the proposed composite loss function~\eqref{eq: NBA loss}. As for RED regularizers, we set by default the denoiser as NLM~\cite{NLM} and $\mu_E=\mu_A=0.1$, $\alpha_1=0.1$, $\alpha_2=0.001$, $\alpha_3=1.0$, $\alpha_4=\alpha_5=0.001$.

\subsubsection{The Effect of Unfolding Network Structure}
\begin{figure*}[!htb]
\centering
\subfloat[\footnotesize RMSE versus epoch]{\includegraphics[width=0.3\linewidth]{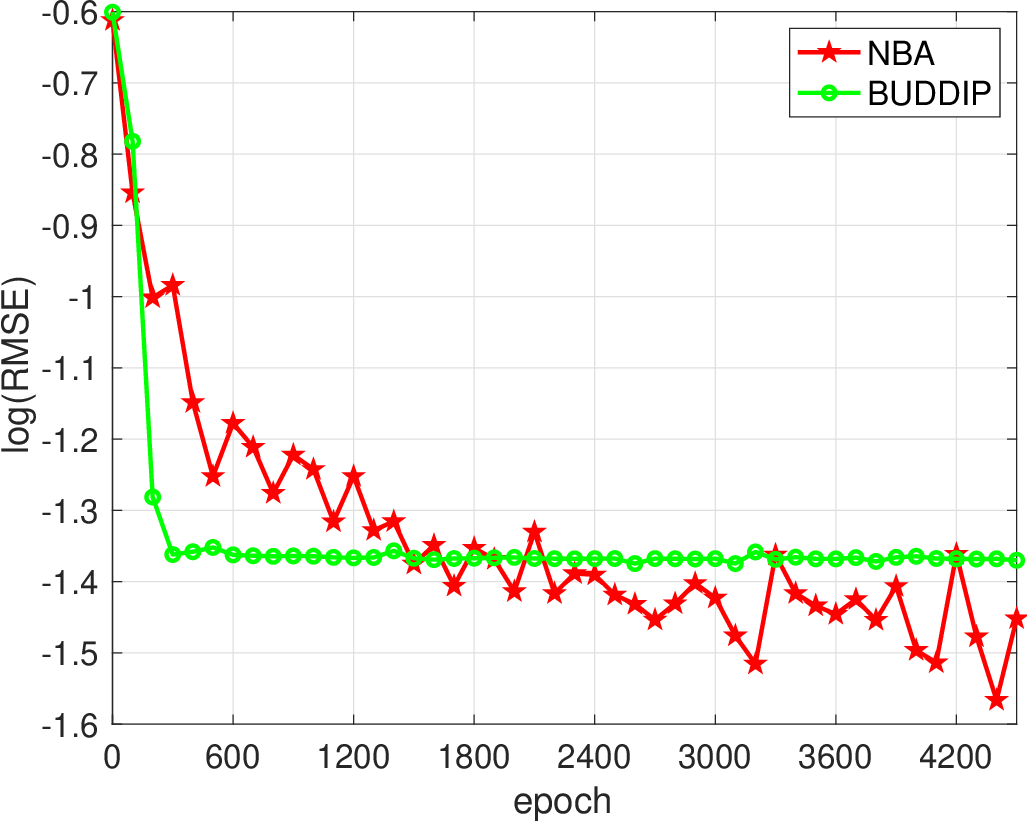}%
\label{fig: synthetic_unfolding_BU_RMSE_vs_epoch}}
\hfil
\subfloat[\footnotesize AAD versus epoch]{\includegraphics[width=0.3\linewidth]{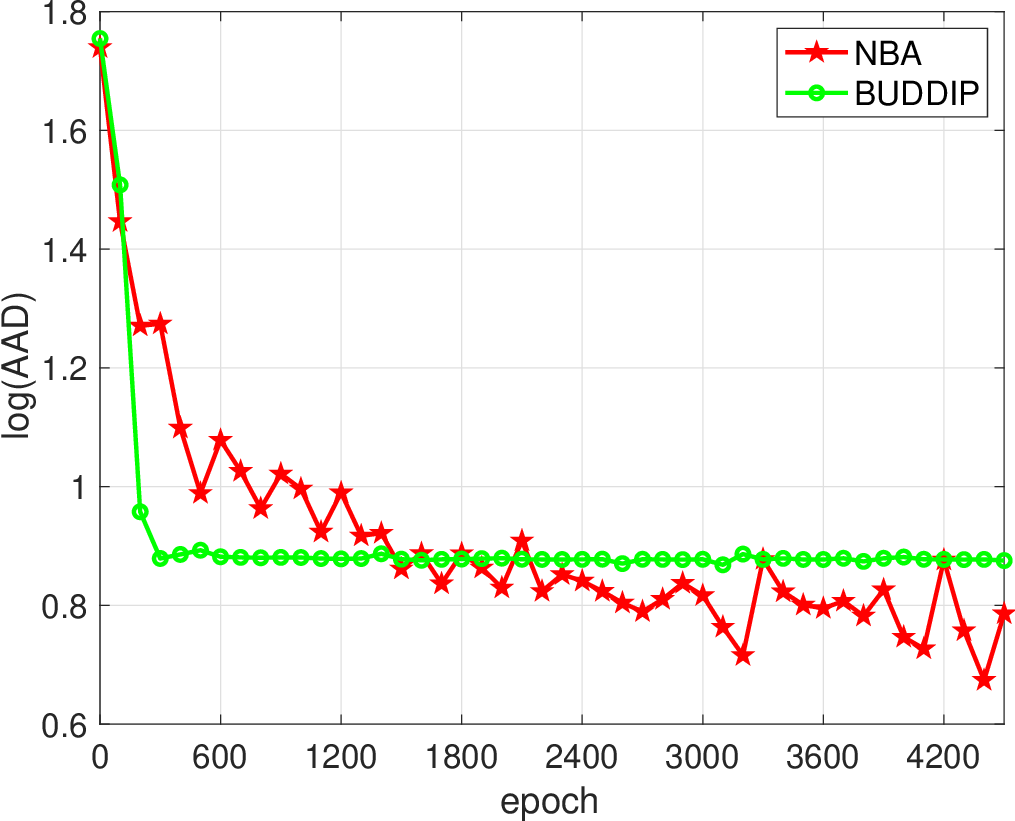}%
\label{fig: synthetic_unfolding_BU_AAD_vs_epoch}}
\hfil
\subfloat[\footnotesize SAD versus epoch]{\includegraphics[width=0.3\linewidth]{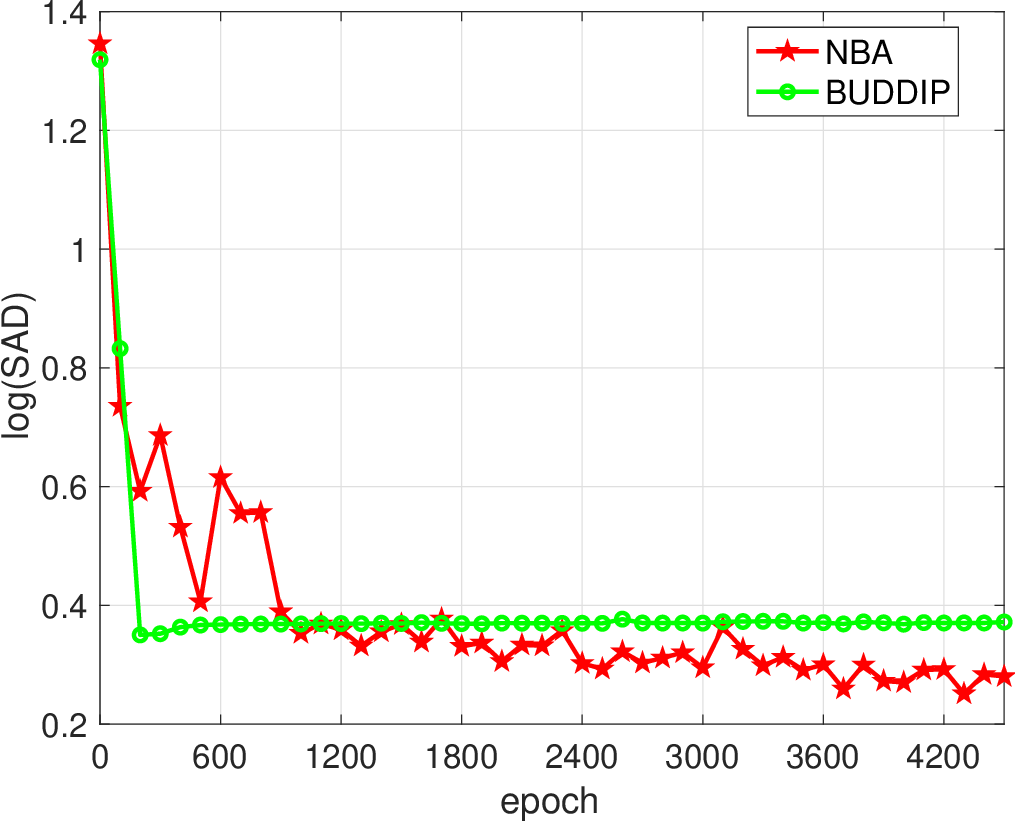}%
\label{fig: synthetic_unfolding_BU_SAD_vs_epoch}}
\caption{The impact of unfolding on blind unmixing performance. The metrics are drawn with the corresponding log value. (a) Abundance RMSE versus epoch. (b) Abundance AAD versus epoch. (c) Endmember SAD versus epoch.}
\label{figs: NBA and BUDDIP vs epoch}
\end{figure*}

We now evaluate the effectiveness of the proposed unfolding structure by comparing the blind unmixing performance between the classical CNN based network, BUDDIP, and unfolding based network, NBA, where the training guidance for both BU networks, i.e., $\{\mathbf{E}_G,\mathbf{A}_G\}$ are generated via SiVM+FCLS. The hyper-parameters of NBA are set as the default values. BUDDIP is also trained using ADAM optimizer with the objective~\eqref{eq: NBA loss}. The results are depicted in Fig.~\ref{figs: NBA and BUDDIP vs epoch}. The final performance is summarized in Table.~\ref{tab:Performance comparison with/-out unfolding}. It is clear that although BUDDIP achieves faster convergence, the NBA network perform better than BUDDIP in terms of the quality of predictions. Specifically, the RMSE of abundance estimation drops from 0.042 to 0.027 and the SAD of endmember estimation decreases from 2.24 to 1.783. In addition, NBA only needs $7.89\times10^5$ learnable parameters whereas BUDDIP has $1.75\times10^6$ learnable parameters. In short, compared to the general CNN based structure, the proposed unfolding structure achieves better performance with fewer learnable parameters.

\subsubsection{The Effect of Proposed MCU based approach}
\begin{figure*}[!htb]
\centering
\subfloat[\footnotesize RMSE versus epoch]{\includegraphics[width=0.3\linewidth]{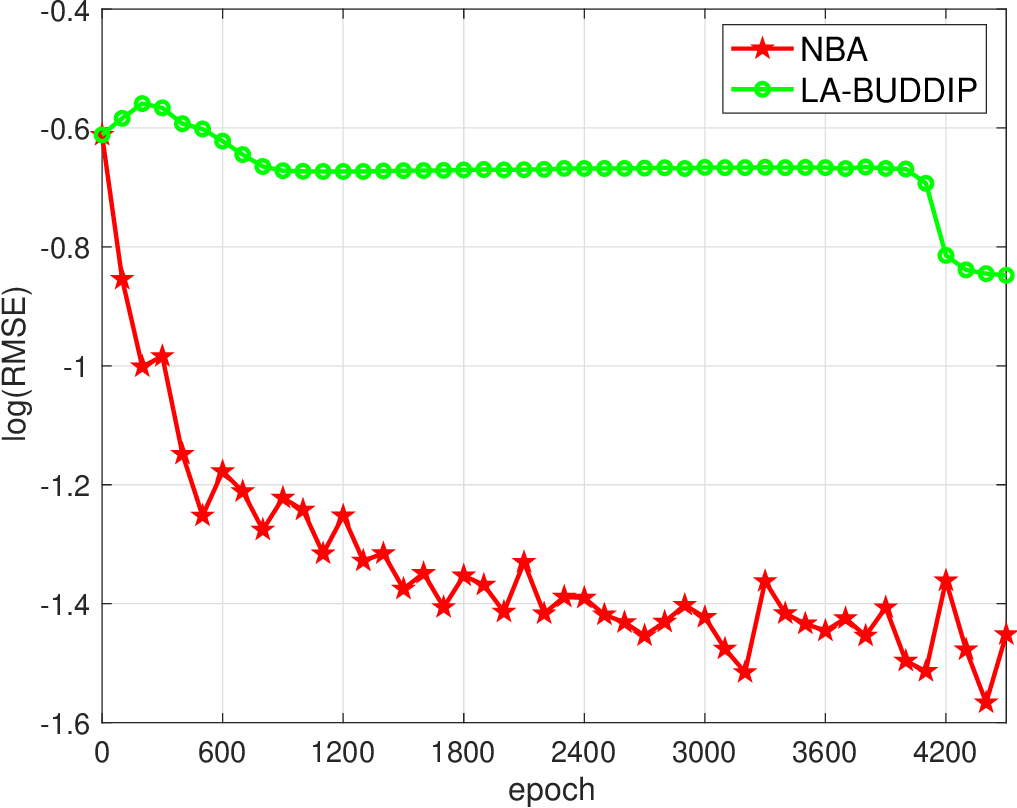}%
\label{fig: syn_MCU_BU_RMSE_vs_epoch}}
\hfil
\subfloat[\footnotesize AAD versus epoch]{\includegraphics[width=0.3\linewidth]{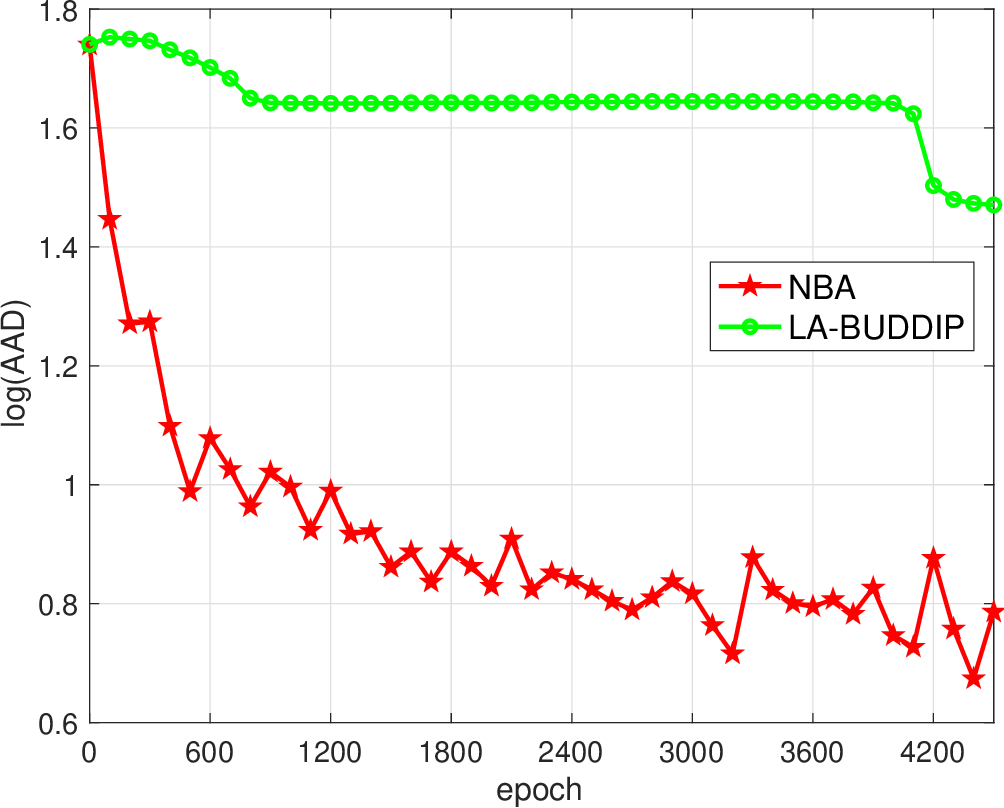}%
\label{fig: syn_MCU_BU_AAD_vs_epoch}}
\hfil
\subfloat[\footnotesize SAD versus epoch]{\includegraphics[width=0.3\linewidth]{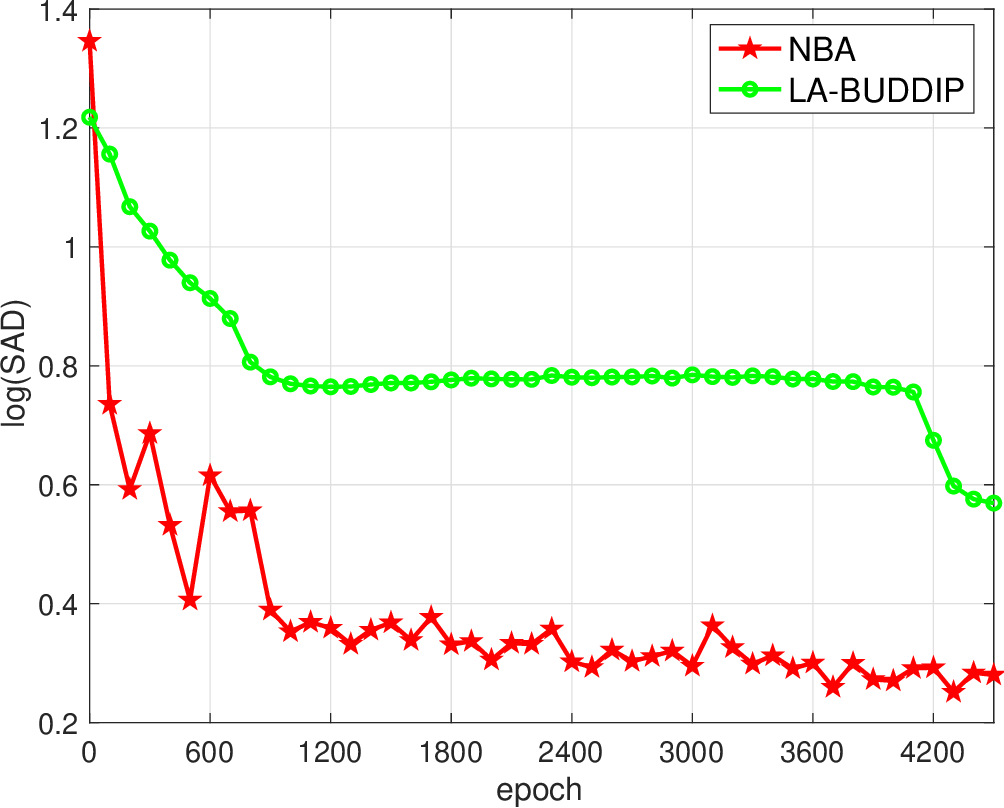}%
\label{fig: syn_MCU_BU_SAD_vs_epoch}}
\caption{Blind unmixing performance comparison between unfolding MCU approach based network (NBA) and unfolding LMM based network (LA-BUDDIP). The metrics are drawn with the corresponding log value. (a) Abundance RMSE versus epoch. (b) Abundance AAD versus epoch. (c) Endmember SAD versus epoch.}
\label{figs: NBA and LA-BUDDIP vs epoch}
\end{figure*}

In this experiment, we evaluate the effectiveness of the proposed MCU unmixing approach by comparing NBA with another BUDDIP structure that is unrolled from the ADMM solver to the linear mixing model. Specifically, we start from the original linear EE model~\eqref{eq: Endmember estimation problem} and linear AE model~\eqref{eq: Abundance estimation problem} and build the corresponding ADMM solvers. We then apply the same unfolding method to these ADMM solvers to build the BUDDIP unrolled from LMM based ADMM solver, which we coin as LA-BUDDIP. We compare the blind unmixing performance between LA-BUDDIP and NBA, where the training guidance for both BU networks, i.e., $\{\mathbf{E}_G,\mathbf{A}_G\}$ are generated via SiVM+FCLS. The hyper-parameters of NBA are set as the default values. LA-BUDDIP is constructed such that the number of learnable parameters is at the same level as NBA. We also train LA-BUDDIP with ADAM optimizer which minimizes the loss function~\eqref{eq: NBA loss}. The results are depicted in Fig.~\ref{figs: NBA and LA-BUDDIP vs epoch}. The final performance is summarized in Table.~\ref{tab:Performance comparison with/-out unfolding}. It is clear that NBA has better performance and faster convergence than LA-BUDDIP. Specifically, the SAD of endmember estimation for NBA is around 1.78 while that for LA-BUDDIP is 3.71. The RMSE of abundance estimation for NBA and LA-BUDDIP is around 0.027 and 0.14, respectively. We attribute this improvement to the network architecture derived from the MCU unmixing approach.
\begin{table*}[!htb]
\begin{center}
\renewcommand{\arraystretch}{1.3}
\caption{Comparison between normal DIP networks and unfolding based DIP networks.}
\label{tab:Performance comparison with/-out unfolding}
\begin{tabular}{| c| c | c | c | c | c|}
\hline
 \multirow{2}{*}{unfolding approach} & \multirow{2}{*}{network} & \#learnable  & Abundance estimation & Abundance estimation& Endmember estimation\\
    & &parameters& metric: RMSE & metric: AAD & metric: SAD \\
\hline
-&BUDDIP& $1.75\times10^6$ & 0.0420 & 7.3847 & 2.240\\\hline
LMM&LA-BUDDIP& $\mathbf{7.40\times10^5}$ & 0.1420 & 29.5476 & 3.708\\\hline
MCU&NBA & $7.89\times10^5$ & $\mathbf{0.0271}$ & $\mathbf{4.7193}$ & $\mathbf{1.783}$\\\hline
\end{tabular}
\end{center}
\end{table*}

\subsubsection{The Effect of Explicit Regualrizer}
\begin{figure*}[!htb]
\centering
\subfloat[\footnotesize Abundance RMSE versus epoch]{\includegraphics[width=0.3\linewidth]{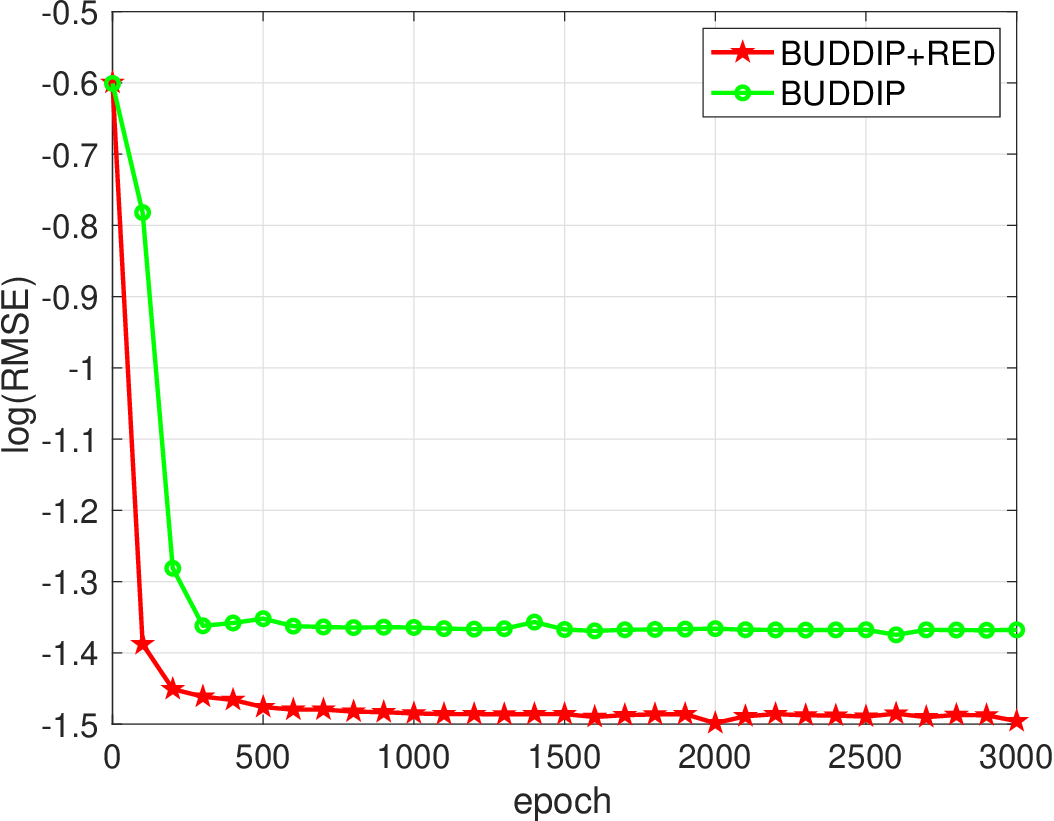}%
\label{fig:syn_RED_BU_RMSE_vs_epoch}}
\hfil
\subfloat[\footnotesize Abundance AAD versus epoch]{\includegraphics[width=0.3\linewidth]{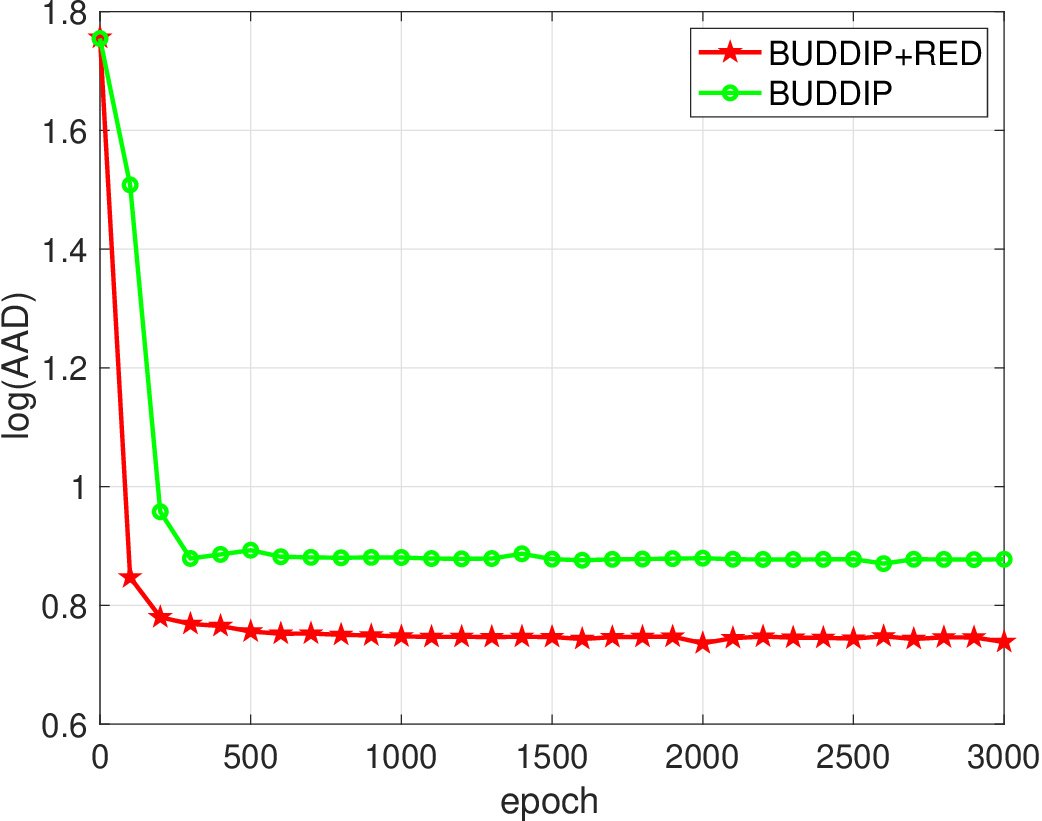}%
\label{fig: syn_RED_BU_AAD_vs_epoch}}
\hfil
\subfloat[\footnotesize Endmember SAD versus epoch]{\includegraphics[width=0.3\linewidth]{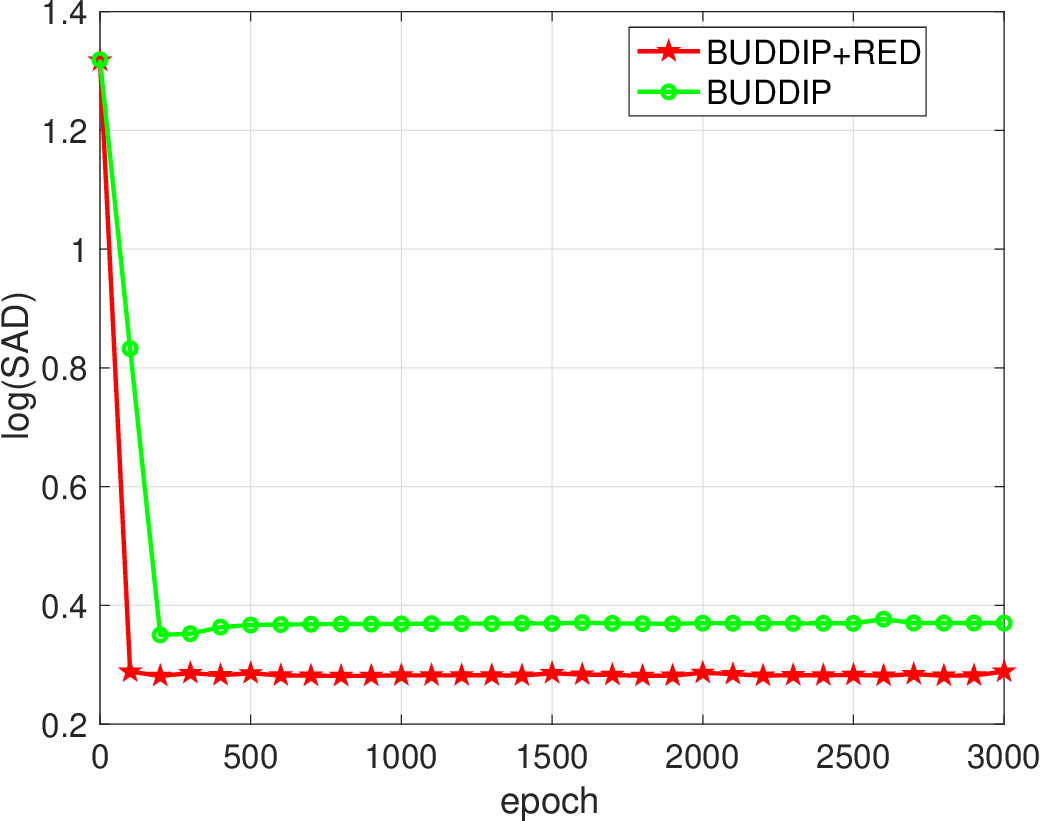}%
\label{fig: syn_RED_BU_SAD_vs_epoch}}
\caption{The impact of RED on blind unmixing performance. The metrics are drawn with the corresponding log value. (a) Abundance RMSE versus epoch. (b) Abundance AAD versus epoch. (c) Endmember SAD versus epoch.}
\label{figs: BUDDIPRED and BUDDIP vs epoch}
\end{figure*}

\begin{table*}[!htb]
\begin{center}
\renewcommand{\arraystretch}{1.3}
\caption{Comparison between DIP networks with and without RED regularizer.}
\label{tab:Performance comparison with/-out RED}
\begin{tabular}{|c |c | c | c | c|}
\hline
 \multicolumn{2}{|c|}{\multirow{2}{*}{network}} & Abundance estimation & Abundance estimation& Endmember estimation\\
  \multicolumn{2}{|c|}{} & metric: RMSE & metric: AAD & metric: SAD \\
\hline
\multirow{3}{*}{BU task}&BUDDIP&  0.0420 & 7.3847 & 2.240\\\cline{2-5}
&BUDDIP+RED & 0.0317 & 5.4470 & 1.889\\\cline{2-5}
&NBA+RED & $\mathbf{0.0257}$ & $\mathbf{4.4496}$ & $\mathbf{1.659}$\\\hline
\end{tabular}
\end{center}
\end{table*}
In this experiment, we present experiments that show the impact of explicit regularizers via comparing blind unmixing performance between BUDDIP and BUDDIP+RED. Specifically, we set the hyperparameters of two RED regularizers as the default settings. The results are reported in Table.~\ref{tab:Performance comparison with/-out RED}. To better show the performance boost, we show the metrics with their corresponding log values in Fig.~\ref{figs: BUDDIPRED and BUDDIP vs epoch}. It is clear that the BUDDIP network with RED regularizers can improve the blind unmixing performance significantly. For example, the SAD of the endmember estimation has improved from 2.24 to 1.889 and the AAD of the abundance estimation has boosted from 7.385 to 5.447. We attribute this improvement to the explicit regularizers added to DIP networks.

\subsubsection{Combing All}
\begin{figure*}[!htb]
\centering
\subfloat[\footnotesize Abundance RMSE versus epoch]{\includegraphics[width=0.3\linewidth]{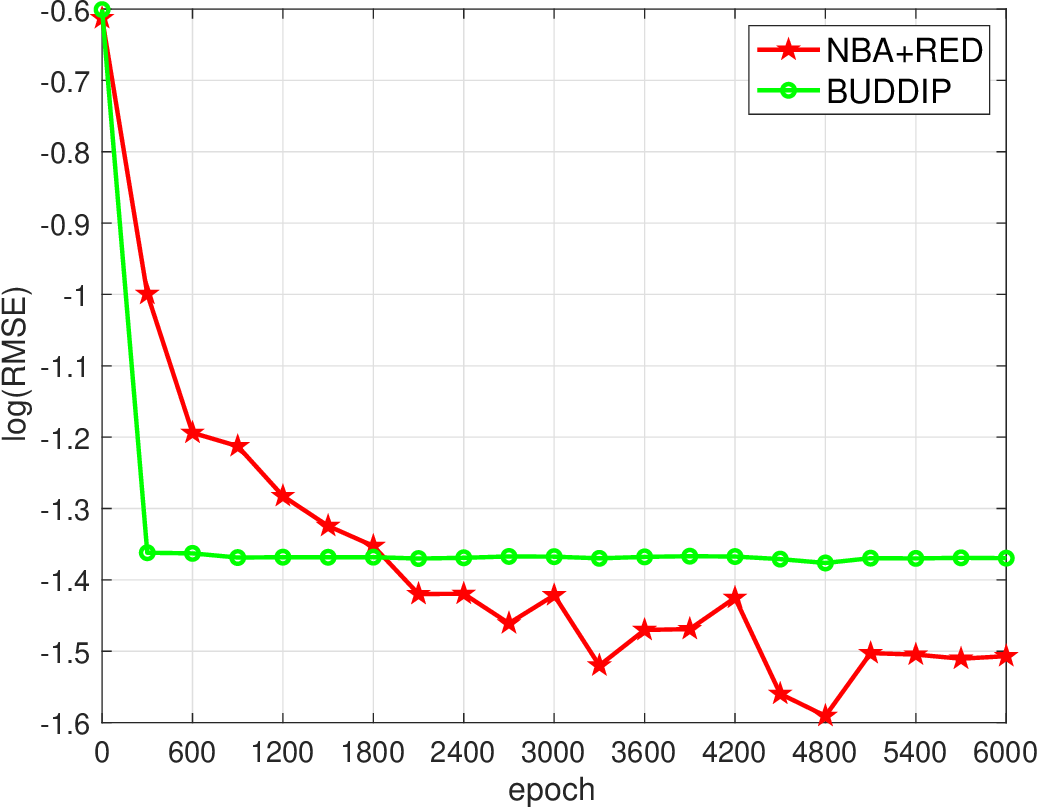}%
\label{fig:syn_unfoldRED_BU_RMSE_vs_epoch}}
\hfil
\subfloat[\footnotesize Abundance AAD versus epoch]{\includegraphics[width=0.3\linewidth]{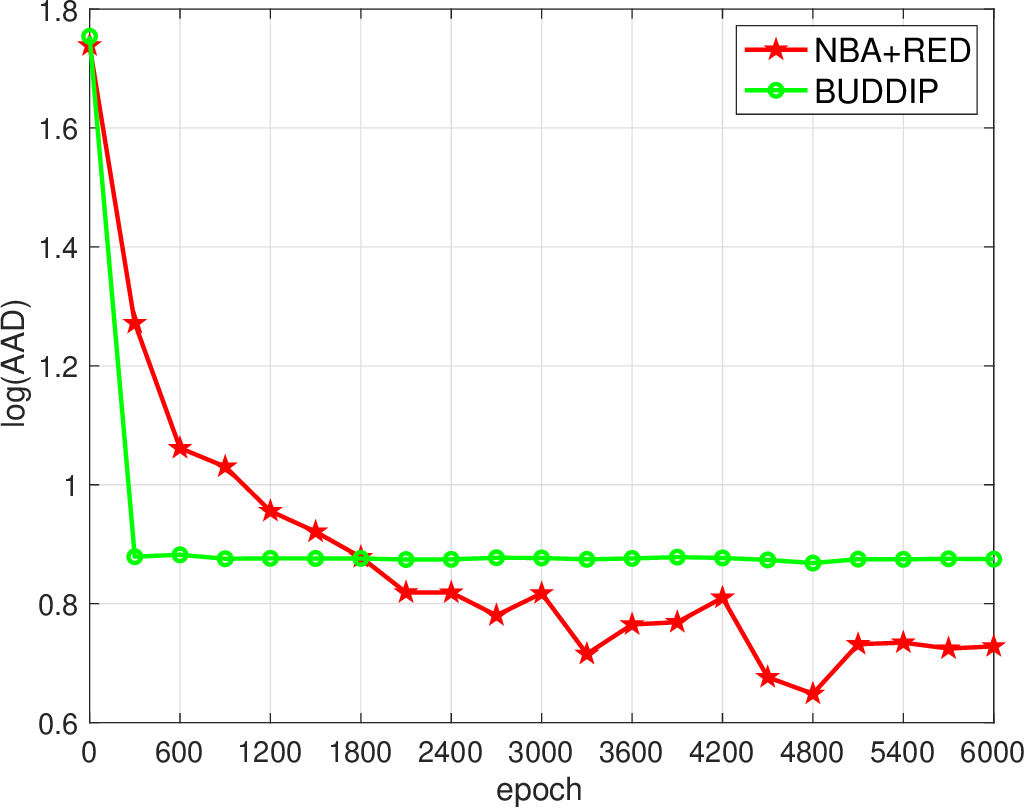}%
\label{fig: syn_unfoldRED_BU_AAD_vs_epoch}}
\hfil
\subfloat[\footnotesize Endmember SAD versus epoch]{\includegraphics[width=0.3\linewidth]{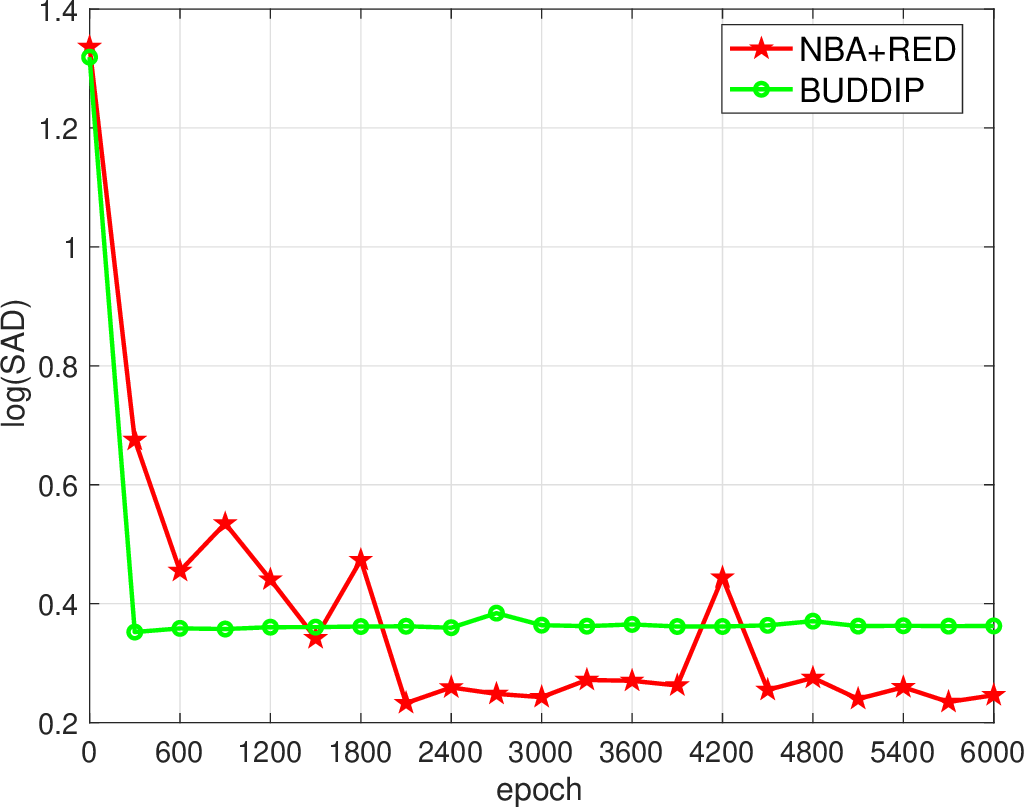}%
\label{fig: syn_unfoldRED_BU_SAD_vs_epoch}}
\caption{Comparison of blind unmixing performance between BUDDIP and NBA+RED. The metrics are drawn with the corresponding log value. (a) Abundance RMSE versus epoch. (b) Abundance AAD versus epoch. (c) Endmember SAD versus epoch.}
\label{figs: BUDDIP and NBARED vs epoch}
\end{figure*}

In this section, we show the overall effectiveness of the proposed methods by comparing the blind unmixing performance between the original BUDDIP and the NBA network enhanced by explicit RED regularizers. The results are reported in Fig.~\ref{figs: BUDDIP and NBARED vs epoch} and Table.~\ref{tab:Performance comparison with/-out RED}. It is clear that BUDDIP provides faster convergence and delivers AE with RMSE$=0.042$ and AAD$=7.38$, and EE with SAD$=2.24$. In comparison, the proposed NBA+RED shows the best blind unmixing performance, i.e., the RMSE and AAD of AE are 0.026 and 4.450, and the SAD of EE is 1.66. 

\subsubsection{Hyperparameter Tuning}
\begin{table}[!htb]
\begin{center}
\renewcommand{\arraystretch}{1.3}
\caption{Hyperparameter tuning.}
\label{tab: hyper param tune}
\begin{tabular}{| c| c| c| c| c| c|}
\hline
$\alpha_1$ & $\alpha_2$ & $\alpha_3$ & RMSE & AAD & SAD\\ \hline
0.001 & 0.001 & 1 & 0.2442 & 54.9031 & 21.1657\\\hline
0.001 & 0.01 & 1 & 0.2387 & 53.3415 &	22.3389\\\hline
0.001 & 0.1 & 1 & 0.23 &	50.4799 &	23.9072\\\hline
0.01 & 0.001 & 1 & 0.2382 &	52.9666 &	22.7712\\\hline
0.01 & 0.01 & 1 & 0.2285 &	50.1863 &	20.7052\\\hline
0.01 & 0.1 & 1 & 0.2256 &	48.9509 &	22.4393\\\hline
0.1 & 0.001 & 1 & $\mathbf{0.0257}$ &	$\mathbf{4.4496}$ &	$\mathbf{1.6588}$\\\hline
0.1 & 0.01 & 1 & 0.0268	& 4.5942 &	1.7046\\\hline
0.1 & 0.1 & 1 & 0.2075 &	43.7296 &	19.7209\\\hline
\end{tabular}
\end{center}
\end{table}
As the hyperparameters $\alpha_1,\alpha_2, \alpha_3$ controls the relative importance of each loss term, it is also necessary to understand how these values affect the performance of NBA+RED. We retain the default settings except that the hyperparameters $\alpha_3$ is fixed to be $1$ and $\alpha_1,\alpha_2$ now vary in the range $[0.1,0.01,0.001]$. The results are reported in Table.~\ref{tab: hyper param tune}. It is clear that with $\alpha_1=0.1, \alpha_2=0.001$, the proposed method delivers the best unmixing performance on the synthetic dataset.
\begin{figure*}[!htb]
\centering
\subfloat[\footnotesize RMSE versus SNR]{\includegraphics[width=0.3\linewidth]{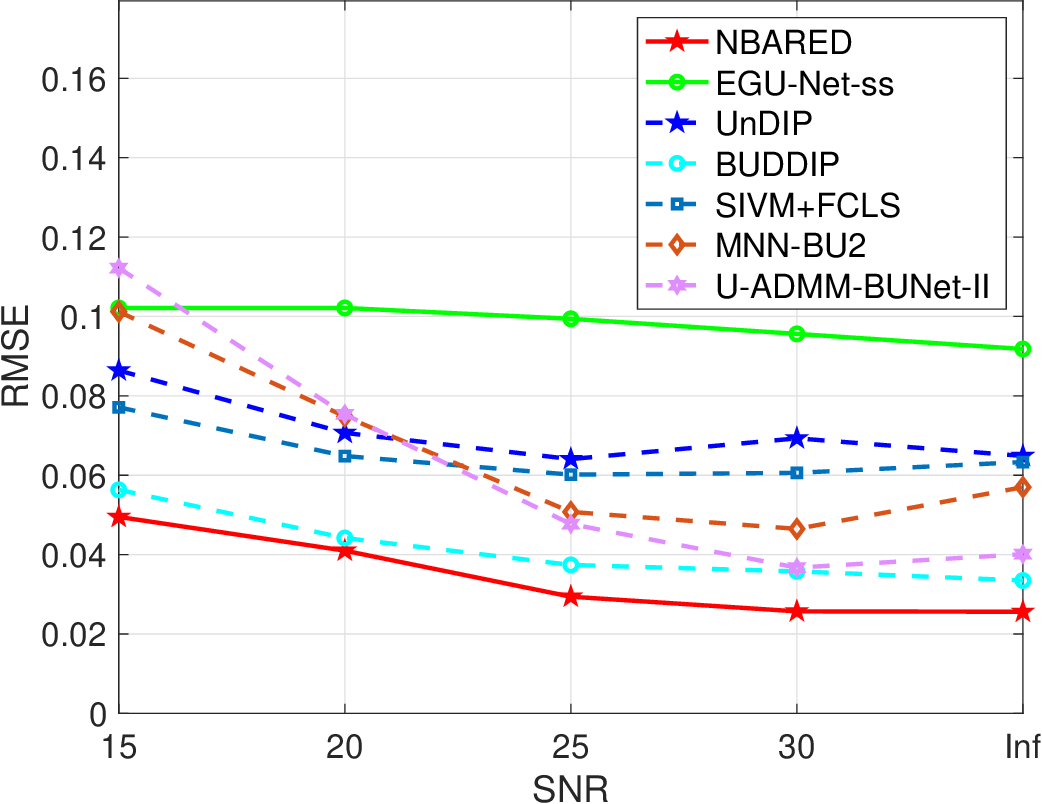}%
\label{fig: RMSE_vs_SNR}}
\hfil
\subfloat[\footnotesize AAD versus SNR]{\includegraphics[width=0.3\linewidth]{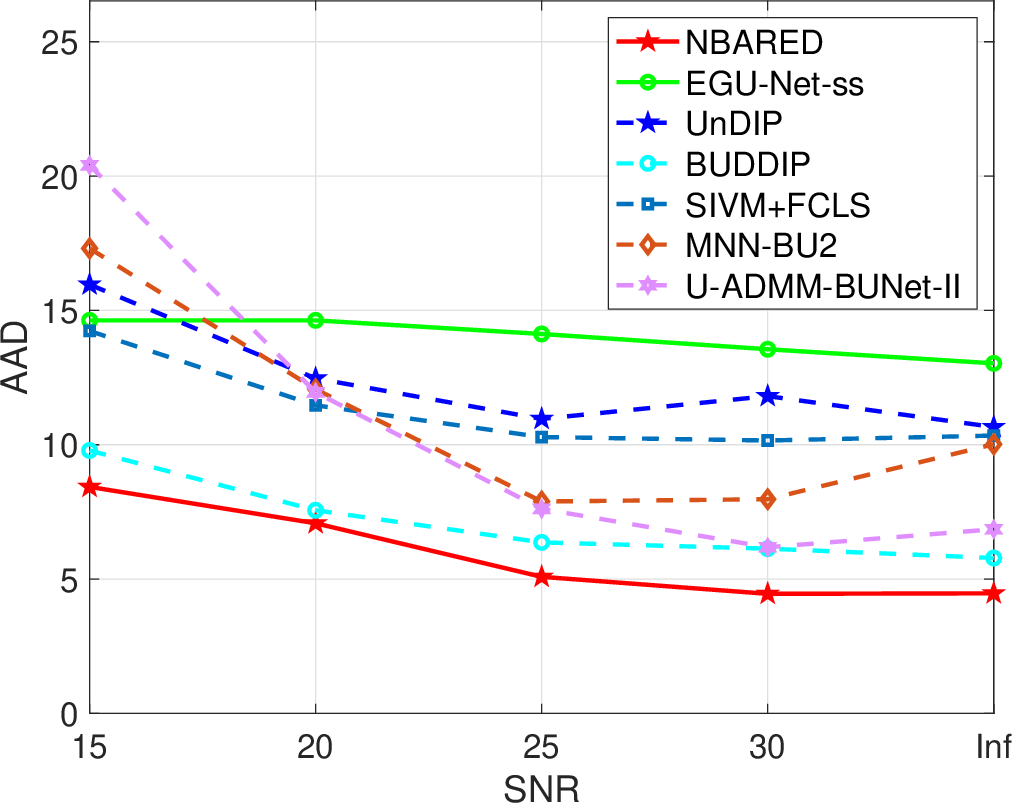}%
\label{fig: AAD_vs_SNR}}
\hfil
\subfloat[\footnotesize SAD versus epoch]{\includegraphics[width=0.3\linewidth]{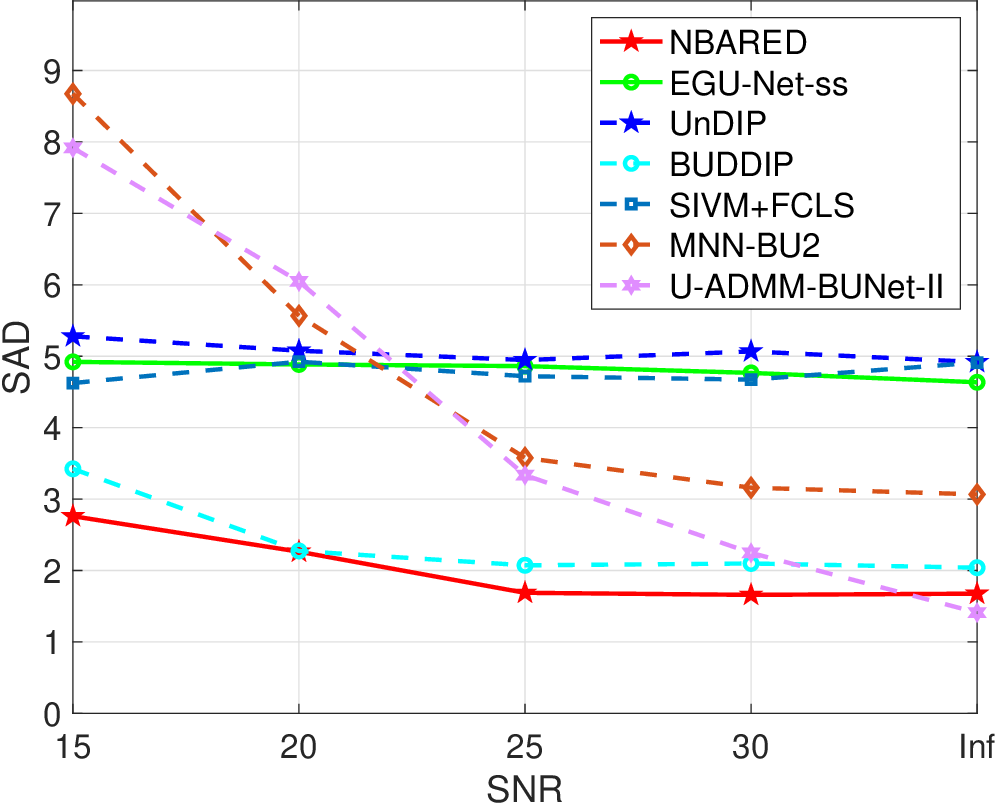}%
\label{fig: SAD_vs_SNR}}
\caption{Blind unmixing performance of various methods versus SNR (dB). (a) Abundance RMSE versus SNR. (b) Abundance AAD versus SNR. (c) Endmember SAD versus SNR.}
\label{figs: performance comparison vs SNR}
\end{figure*}

\subsection{Evaluation on Synthetic Data}
We now evaluate the effectiveness of the proposed approach NBARED on the synthetic dataset, in comparison with some state-of-the-art approaches including, MNN-BU, EGU-Net, UADMM-BUNet, and UnDIP. We also compare with the traditional method SiVM~\cite{Heylen2011}+FCLS~\cite{Heinz2001} which is also used to generate the training guidance. We also compare it with BUDDIP~\cite{Zhou2022}. We use the same default settings for the synthetic HSI image and NBAsRED network in Section.~\ref{sec: ablation study}. Specifically, we test the robustness of various methods against AWGN noise. We retain the default settings except that the SNR of synthetic data varies in $[15,20,25,30,\infty]$ dB. It is worth mentioning that MNN-BU and UADMM-BUNet, similar to UnDIP and EGU-Net, require existing methods to generate an estimation of endmembers and abundances to initialize the network. In this paper, the initialisation is also generated by SiVM+FCLS for a fair comparison. The results are illustrated in Fig.~\ref{figs: performance comparison vs SNR}. It is clear that all methods enjoy benefits from increased SNR. It is also noticeable that other training guidance based methods such as UnDIP and EGU-Net have the drawback that their performance is limited by the quality of guidance generated by SiVM+FCLS. On the other hand, UADMM-BUNet, MNN-BU, BUDDIP and NBARED does not suffer from this limitation and shows better performance than the guidance SiVM+FCLS. When SNR is small, the doubleDIP structure based BUDDIP and NBARED achieves better performance than UADMM-BUNet and MNN-BU. This performance gain is smaller as the SNR increases. Although BUDDIP already shows significant better robustness against AWGN noise than other state-of-art methods. The proposed NBARED achieves even better performance than BUDDIP. 


\subsection{Evaluation on Real Data}
\subsubsection{Jasper Ridge}
\begin{figure*}[!htb]
\centering
\subfloat[\scriptsize SiVM+FCLS]{\includegraphics{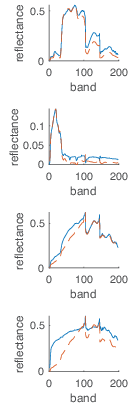}%
\label{fig: SiVM+FCLS endm}}
\hfil
\hspace{-1em}
\subfloat[\scriptsize MNN-BU-2]{\includegraphics{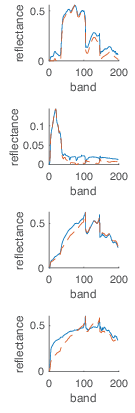}%
\label{fig: MNNBU endm}}
\hfil
\hspace{-1em}
\subfloat[\scriptsize U-ADMM-\\\centering{BUNet-II}]{\includegraphics{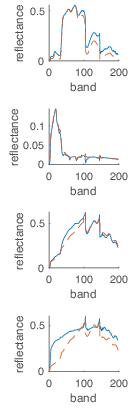}%
\label{fig: UADMMBUNet endm}}
\hfil
\hspace{-1em}
\subfloat[\scriptsize EGU-Net-ss]{\includegraphics{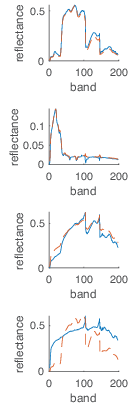}%
\label{fig: EGUNetss endm}}
\hfil
\hspace{-1em}
\subfloat[\scriptsize UnDIP]{\includegraphics{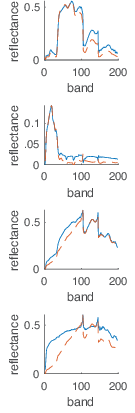}%
\label{fig: UnDIP endm}}
\hfil
\hspace{-1em}
\subfloat[\scriptsize BUDDIP]{\includegraphics{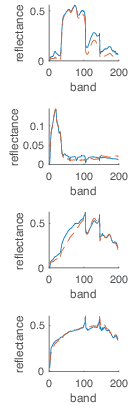}%
\label{fig: BUDDIP endm}}
\hfil
\hspace{-1em}
\subfloat[\scriptsize rNMF]{\includegraphics{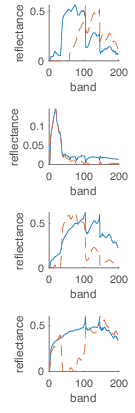}%
\label{fig: rNMF endm}}
\hfil
\hspace{-1em}
\subfloat[\scriptsize NBARED]{\includegraphics{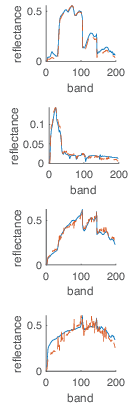}%
\label{fig: NBARED endm}}
\caption{Endmembers estimated by different methods on Jasper Ridge dataset. Blue solid lines indicate the true value, while red dot lines indicate the scaled estimated value. From top to bottom: Tree, Water, Soil, and Road. (a) SiVM+FCLS. (b) MNN-BU-2. (c) U-ADMM-BUNet-II. (d) EGU-Net-ss. (e) UnDIP. (f) BUDDIP. (g) rNMF. (h) NBARED.}
\label{figs: JasperRidge Endm Signature}
\end{figure*}

\begin{figure*}[!htb]
\centering
\subfloat[\scriptsize SiVM+FCLS]{\includegraphics{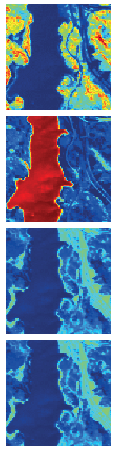}%
\label{fig: SiVM+FCLS  abudnance map}}
\hfil
\hspace{-0.6em}
\subfloat[\scriptsize MNN-BU-2]{\includegraphics{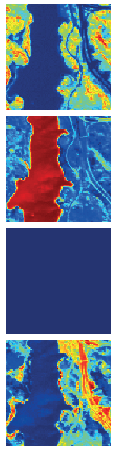}%
\label{fig: MNNBU  abudnance map}}
\hfil
\hspace{-0.6em}
\subfloat[\scriptsize U-ADMM-\\\centering{BUNet-II}]{\includegraphics{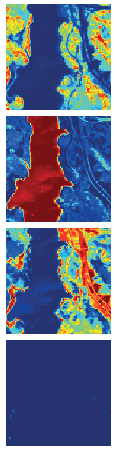}%
\label{fig: UADMMBUNet  abudnance map}}
\hfil
\hspace{-0.6em}
\subfloat[\scriptsize EGU-Net-ss]{\includegraphics{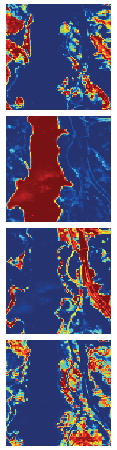}%
\label{fig: EGUNetss  abudnance map}}
\hfil
\hspace{-0.6em}
\subfloat[\scriptsize UnDIP]{\includegraphics{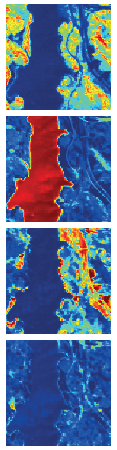}%
\label{fig: UnDIP  abudnance map}}
\hfil
\hspace{-0.6em}
\subfloat[\scriptsize BUDDIP]{\includegraphics{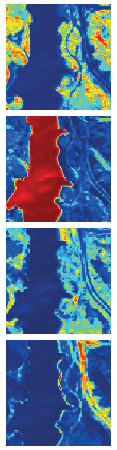}%
\label{fig: BUDDIP abudnance map}}
\hfil
\hspace{-0.6em}
\subfloat[\scriptsize rNMF]{\includegraphics{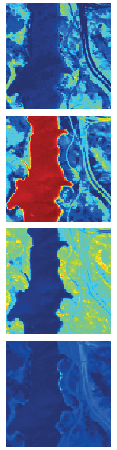}%
\label{fig: rNMF abudnance map}}
\hfil
\hspace{-0.6em}
\subfloat[\scriptsize NBARED]{\includegraphics{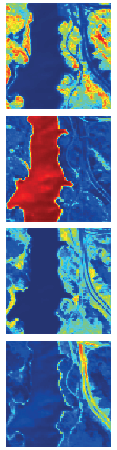}%
\label{fig: NBARED abudnance map}}
\hfil
\hspace{-0.6em}
\subfloat[\scriptsize Reference]{\includegraphics{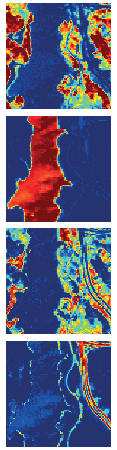}%
\label{fig: Jasper Ridge abudnance ref}}
\hfil
\hspace{-0.6em}
\subfloat{\includegraphics{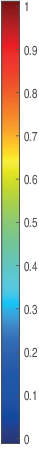}%
}
\caption{Abundance maps estimated by different methods on Jasper Ridge dataset. From top to bottom: Tree, Water, Soil, and Road. (a) SiVM+FCLS. (b) MNN-BU-2. (c) U-ADMM-BUNet-II. (d) EGU-Net-ss. (e) UnDIP. (f) BUDDIP. (g) rNMF. (h) NBARED. (i) Reference.}
\label{figs: JasperRidge abundance maps}
\end{figure*}

We now evaluate the effectiveness of the proposed method NBARED on real data Jasper Ridge, by comparing it with some of the state-of-the-art methods, including the traditional methods SiVM+FCLS and rNMF, learning based methods MNN-BU, UADMM-BUNet, EGU-Net, UnDIP and BUDDIP~\cite{Zhou2022}. The hyperparameters of the proposed NBARED are set to $J_E=1, J_A=3, m_E=m_A=128, k_E=k_A=5, \alpha_1=\alpha_2=\alpha_3=1.0,\alpha_4=0.01, \alpha_5=\mu_E=\mu_A=0.001$, the learning rate and $\beta_2$ of ADAM optimizer is set to $5e-4$ and 0.85, respectively. The qualitative results of estimated endmembers and abundances are illustrated in Fig.~\ref{figs: JasperRidge Endm Signature} and Fig.~\ref{figs: JasperRidge abundance maps}. The corresponding quantitative results are reported in Table.~\ref{table: linear unmixing quantitative comparisons}, where the best one is shown as bold. It can be seen that the proposed NBARED show better accuracy for abundance estimation as it delivers the best RMSE as 0.1124 and the best AAD as 14.0751. As for the endmember estimation, EGU-Net gives the best estimation for the endmember Tree with SAD as 2.76. U-ADMM-BUNet also shows the best estimation for Water with SAD to be 4.09. Nevertheless, BUDDIP yields a SAD as 2.61 in terms of estimation for Road and NBARED achieves the best estimation for Soil with SAD as 3.74. NBARED also achieves better performance than BUDDIP in terms of endmember estimation for Tree, Water and Soil, at the expense that the estimation for Road is worse. On average, the proposed NBARED achieves the best RMSE, AAD for abundance estimation and the best SAD for abundance estimation, in comparison with other state-of-the-art methods.


\begin{table*}[!htb]
\renewcommand{\arraystretch}{1.3}
\caption{Abundance RMSE, AAD (In Degrees), endmember SAD (In Degrees) by Different methods on Jasper Ridge. The Best Results are in Bold.}
\label{table: linear unmixing quantitative comparisons}
\centering
\begin{tabular}{c| c c |c c c c c} \toprule
 Methods     & RMSE & AAD & SAD of Tree & SAD of Water & SAD of Soil & SAD of Road & averaged SAD \\\midrule
SiVM+FCLS & 0.1480    & 20.7198 & 8.5545 & 14.4877 & 6.5558 & 15.7991  & 11.3493  \\ \midrule
MNN-BU-2 & 0.2154   & 33.1936  & 7.9771 & 14.4324 & 5.3219 & 8.6187 & 9.0875\\ \midrule
U-ADMM-Net-II & 0.1332    & 17.6087  & 8.6635 & \textbf{4.0919} & 5.6216 & 13.6586  & 8.0089\\ \midrule
EGU-Net-ss & 0.2110   & 30.7221  & \textbf{2.7626} & 4.5969 & 5.6690 & 22.0731  & 8.7754\\  \midrule
UnDIP & 0.1748    & 25.3249 & 8.5545 & 14.4877 & 6.5558 & 15.7991  & 11.3493    \\ \midrule
BUDDIP & 0.1176  & 14.9592  & 8.7274 & 8.7116 & 6.8511 & \textbf{2.6116}  & 6.7255\\
\midrule
rNMF & 0.2260    & 34.4614& 47.0341 & 16.1319 & 33.3038 & 32.0590 & 32.1322    \\ 
\midrule
NBARED & \textbf{0.1124}    & \textbf{14.0751} & 4.7596 & 5.9893 & \textbf{3.7364} & 8.5702 & \textbf{5.7638}    \\ \bottomrule
\end{tabular}
\end{table*}

\section{Conclusion}
In this work, we have proposed a novel neural network structure for hyperspectral blind unmixing problems. Firstly, different from the general linear mixture model (LMM), we propose a MatrixConv Unmixing (MCU) approach, for which we have also proposed AE and EE recovery problems and the corresponding ADMM solvers. We then apply the algorithm unrolling technique to each solver to construct unfolding based deep image prior networks for endmember estimation  (UEDIP) and abundance estimation (UADIP), respectively. We then combine these two networks based upon LMM to give our final network for blind unmixing based on ADMM solvers, which we coined as NBA. In addition, we also propose to add explicitly two regularizers by denoising (RED) for endmember estimation and abundance estimation, respectively. We finally propose an ADMM solver for the NBA network enhanced by RED (NBARED). The experiments on both synthetic and real datasets show that the proposed method outperforms the state-of-the-art unmixing methods including MNN-BU, UADMM-BUNet and EGU-Net.

\section*{Acknowledgments}
This should be a simple paragraph before the References to thank those individuals and institutions who have supported your work on this article.



\bibliographystyle{IEEEtran}
\bibliography{bib/IEEEabrv.bib,bib/ref.bib}

\newpage

 




\vfill

\end{document}